\documentclass[preprint,showpacs,showkeys,aps,prc]{revtex4-1}
\usepackage{graphicx}
\usepackage{dcolumn}
\usepackage{bm}%

\begin{document}

\title{Mixed phase in a compact star with strong magnetic field}

\date{\today}

\author{Ritam Mallick\footnote{Email:ritam.mallick5@gmail.com} \& P K Sahu\footnote{Email:pradip@iopb.res.in}}
\affiliation{Institute of Physics, Sachivalaya Marg,  
        Bhubaneswar 751005, INDIA}

\begin{abstract}
Compact stars can have either hadronic matter or can have exotic states of matter
like strange quark matter or color superconducting matter. Stars also can have
a quark core surrounded by hadronic matter, known as hybrid stars (HS). The HS 
is likely to have a mixed phase in between the hadron and quark phase. Observational results 
suggest huge surface magnetic field in certain neutron stars (NS) called magnetars.
Here we study the effect of strong magnetic field on the 
respective EOS of matter under extreme conditions. 
We further study the hadron-quark phase transition in the interiors
of NS giving rise to hybrid stars (HS) in presence of strong magnetic field. The hadronic 
matter EOS is described based on relativistic mean field theory and we include the effect of strong 
magnetic fields leading to Landau quantization of the charged particles. For the 
quark phase we use the simple MIT bag model. 
We assume density dependent bag pressure and magnetic field.
The magnetic field strength increases going from the surface to the center of the star.
We construct the intermediate mixed phase using Glendenning conjecture. 
The magnetic field softens the EOS of both the matter phases.
The effect of magnetic field is insignificant unless the field strength
is above $10^{14}$G. A varying magnetic field, with surface field strength 
of $10^{14}$G and the central field strength of the order of $10^{17}$G 
has significant effect on both the stiffness and the mixed phase regime 
of the EOS. We finally study the mass-radius relationship
for such type of mixed HS, calculating their maximum mass, and compare them with the 
recent observation of pulsar PSR J1614-2230, which is about 2 solar mass. 
The observations puts a severe constraint on the EOS of matter at extreme conditions.
The maximum mass with our EOS can reach the limit set by the observation. 
\end{abstract}



\maketitle

\section*{Introduction}

The central density of neutron stars exceed the nuclear saturation density 
($n_0 \sim 0.15\:$fm$^{-3}$), thereby raising the idea that compact 
stars might contain deconfined and chirally restored quark matter in them. 
Recently, \cite{Demorest10} the mass measurement of millisecond pulsar PSR J1614-2230 
has set a new robust mass limit for compact stars to be $M=1.97 \pm 0.04\:$M$_\odot$. 
This value, together with the mass of pulsar J1903+0327 of $M=1.667 \pm 0.021\:$M$_\odot$ 
\cite{Freire10} is much larger than any of the highest precisely measured pulsar mass. 
These measurement has set for the first time a very strong limit on parameters of the 
EOS, which describes matter under extreme conditions \cite{weber, glendenning}. 

After the discovery of pulsar \cite{hewish} and connecting them with NS \cite{gold},  
various EOS for nuclear matter has been proposed and refined \cite{bethe, walecka,
baldo, akmal}. The quark sector is not much well understood as the nature of strong
interaction at extreme condition still remains a challenge. 
The strange quark matter (SQM) conjecture by Itoh, Witten \cite{itoh,witten}
consisting of almost equal number of up (u), down (d) and strange (s) quarks was 
supported by model calculations \cite{alcock}. The most simple and popular model which
describes the properties of quark matter at such high densities is the 
MIT bag model \cite{chodos}. New refined models based 
on results from recent experiments in laboratories has been proposed \cite{fowler,dey,zhang}. 
Thus normal nuclear matter at high density and/or 
temperature is likely to be unstable against stable SQM and would eventually decay. 

Compact objects therefore can be made of either nuclear matter or quark matter.
Stars which has only nuclear matter are called neutron stars (NS).
Broadly there can be two classes of compact stars with quark matter. 
The first is the so-called (strange) quark stars (SS) of absolutely stable strange 
quark matter. The second are the so-called hybrid stars (HS), along with the hadronic matter they have 
quarks matter in their interior 
either in form of a pure strange quark matter core or color superconducting matter. In between the quark 
and the hadronic phase a quark-hadron mixed phase exists. 
The size of the core depends on the critical density for the quark-hadron phase 
transition and the EOS describing the matter phases. 

Usually, the presence of strangeness in quark and hadronic matter provides an additional 
degree of freedom and softens the EOS and therefore quark and hybrid stars cannot reach 
high masses. Thus the mass measurement of pulsar PSR J1614-2230 puts forward a strong 
constraint on such EOS. However, studies found that effects from the strong interaction, such as 
one-gluon exchange or color-superconductivity can stiffen the quark matter EOS and increase 
their maximum mass \cite{Ruester04,Horvath04,Alford07,Fischer10,Kurkela10a,Kurkela10b}.
The first studies on the implications of the new mass limits from 
PSR J1614-2230 for quark matter was done by \cite{Ozel10} and \cite{Lattimer10}. 
They, however did not include the effects from color-superconductivity. 

The presence of magnetic field in compact stars has an important role in astrophysics.
New observations suggests that in some pulsars the surface magnetic field can be as high 
as  $10^{14}-10^{15}$G. It has also been attributed that the observed giant flares, 
SGR 0526-66, SGR 1900+14 and SGR 1806-20 \cite{flare}, are the manifestation of such strong 
surface magnetic in those stars. Such stars are separately assigned as magnetars. If
we assume flux conservation from a progenitor star, we can expect the central 
magnetic field of such stars as high as $10^{17}-10^{18}$G. Such strong fields are 
bound to effect the NS properties. It can modify the metric describing the star  
\cite{bocquet,cardall} or it can modify the EOS of matter of the star. 
The effect of strong magnetic field, both for nuclear matter 
\cite{chakrabarty97,yuan,broderick,chen,wei} and quark matter 
\cite{chakrabarty96,ghosh,felipe} has been studied earlier in detail.

Motivated by recent observations of maximum mass limits of compact stars and strong magnetic
field in magnetars, in this work we want to explore their implications on the EOS of both phases 
of matter that may be present inside a neutron star. We study the hadron-quark phase transition
inside a compact star with a mixed phase region in between the quark core and nuclear outer region.
The paper is organized as follows. In Sec. II we discuss the nuclear EOS and the effect of Landau
quantization due to magnetic field on the charged particles. In Sec. III we employ the simple MIT 
bag model for the quark matter EOS and the effect of magnetic field on the quarks (also due to Landau 
quantization). In Section IV we develop the mixed phase region by Glendenning construction. 
We show our results in section V for the density dependent bag constant and 
varying magnetic field for the mixed HS. Finally we summarize our results and draw some conclusion 
in section VI.

\section*{Magnetic field in hadronic phase}

At normal nuclear density the degrees of freedom for the EOS are hadrons. To describe the hadronic phase, 
we use a non-linear version of the relativistic mean field (RMF) model with hyperons (TM1 parametrization) 
which is widely used to 
construct EOS for NS. In this model the baryons interact with mean  
meson fields \cite{boguta,glen91,sugahara,sghosh,schaffner,schertler}.

The Lagrangian density including nucleons, baryon octet ($\Lambda,\Sigma^{0,\pm},\Xi^{0,-}$) 
and leptons is given by 
\begin{eqnarray} 
\label{baryon-lag}   
{\cal L}_H & = & \sum_{b} \bar{\psi}_{b}[\gamma_{\mu}(i\partial^{\mu}  - g_{\omega b}\omega^{\mu} - 
\frac{1}{2} g_{\rho b}\vec \tau . \vec \rho^{\mu})  \nonumber \\ 
& - & \left( m_{b} - g_{\sigma b}\sigma \right)]\psi_{b} + \frac{1}{2}({\partial_\mu \sigma \partial^\mu 
\sigma - m_{\sigma}^2 \sigma^2 } ) \nonumber \\ 
& - & \frac{1}{4} \omega_{\mu \nu}\omega^{\mu \nu}+ \frac{1}{2} m_{\omega}^2 \omega_\mu \omega^\mu - 
\frac{1}{4} \vec \rho_{\mu \nu}.\vec \rho^{\mu \nu} \nonumber \\
& + & \frac{1}{2} m_\rho^2 \vec \rho_{\mu}. \vec \rho^{\mu} -\frac{1}{3}bm_{n}(g_{\sigma}\sigma)^{3}-
\frac{1}{4}c(g_{\sigma}\sigma)^{4} +\frac{1}{4}d(\omega_{\mu}\omega^{\mu})^2 \nonumber \\
& + & \sum_{L} \bar{\psi}_{L}    [ i \gamma_{\mu}  \partial^{\mu}  - m_{L} ]\psi_{L}.
\end{eqnarray}
Leptons ${\cal L}$ are treated as non-interacting and baryons $b$ are coupled to the scalar meson $\sigma$, 
the isoscalar-vector meson $\omega_\mu$ and the isovector-vector meson $\rho_\mu$. 
There are five constants in the model that are fitted to the bulk properties of nuclear matter. 
This model is good enough to describe nuclear matter and the nuclear 
saturation point. But it is insufficient for the hyperonic matter,
because the model does not reproduce the observed strong 
$\Lambda \Lambda$ 
attraction. This defect can be remedied by adding two new meson
fields with hidden strangeness,  namely, the iso-scalar scalar $\sigma^*$ and 
the iso-vector vector $\phi$, which couple to hyperons only \cite{schaffner}.

The effective baryon mass is given by 
\begin{eqnarray}
 {m_b}^*=m_b-g_{\sigma}\sigma-g_{\sigma^*}\sigma^*.
\end{eqnarray}
For the beta equilibrated matter the conditions is
\begin{equation}
\mu_i= b_i\mu_B+q_i\mu_e ,
\label{chm}
\end{equation}
where $b_i$ and $q_i$ are the baryon number and charge (in terms of electron charge) of species $i$, 
respectively.
$\mu_B$ is the baryon chemical potential and $\mu_e$ is the electron chemical potential.
For charge neutrality, the condition is 
\begin{equation}
\rho_c=\sum_i q_i n_i,
\end{equation}
$n_i$ is the baryon number density of particle $i$.

The magnetic field is assumed to be in the $z$ direction, $\overrightarrow{B}=B\hat{k}$.
Now the motion of the charged particles are quantized in the 
perpendicular direction of the magnetic field. The landau quantized energy 
is given by \cite{lan}
\begin{equation}
E_i=\sqrt{{p_i}^2+{m_i}^2+|q_i|B(2n+s+1)}. 
\end{equation}
In the above equation $n$ is the principle quantum number, $s$ is the spin of the 
particle (either (+) or (-)) and $p_i$ is the momentum component along the field direction 
of particle $i$. We can write $2n+s+1=2\nu$, where
$\nu = 0, 1, 2...,$ so that now the energy can be written as 
\begin{equation}
E_i=\sqrt{{p_i}^2+{m_i}^{2}+2\nu |q_i|B} \nonumber \\
=\sqrt{{p_i}^2+{\widetilde{m}_{i,\nu}}^2}  
\end{equation}
where the $\nu=0$ state is singly degenerate. It should be remembered that for 
baryons the mass is ${m_b}^*$.

At zero temperature and in the presence of a constant magnetic field $B$,
the number and energy densities of charged particles are given by \cite{chakrabarty96,broderick}
\begin{equation}
n_i= \frac{|q_i| B}{2 \pi^2} \sum_{\nu}
p_{f,\nu}^i \,,
\label{nmax}
\end{equation}
and 
\begin{equation}
\varepsilon_i= \frac{|q_i| B}{4 \pi^2} \sum_{\nu}
\left[ E_f^i p_{f,\nu}^i + \widetilde{m}_{\nu}^{i~2} \ln
\left( \left|
\frac{E_f^i + p_{f,\nu}^i}{\widetilde{m}^i_{\nu}} \right|
\right) \right] \,.
\end{equation}
$p_{f,\nu}^i$ is the Fermi momentum for the level with the
principal quantum number $n$ and spin $s$ and is given by
\begin{equation}
p_{f,\nu}^{i~2} =  E_f^{i~2} - \widetilde{m}_{\nu}^{i~2} \,.
\end{equation}
The Fermi energies are fixed by their respective chemical potentials.

The number, energy, and scalar number densities of the neutral particles are
given by
\begin{equation}
n_N = \frac{p_f^{N~3}}{3 \pi^2} \,,
\end{equation}
\begin{equation}
n^s_N = \frac{ m_N^*}{2 \pi^2} \left[
E_f^N p_f^N - m_N^{*~2} \ln \left( \left|
\frac{E_f^N + p_f^N}{m_N^*} \right| \right) \right] \,,
\end{equation}
\begin{equation}
\varepsilon_N = \frac{1}{8 \pi^2} \left[ 2 E_f^{N~3} p_f^N - m_N^{*~2}
E_f^N p_f^N -m_N^{*~4} \ln \left( \left| \frac{E_f^N + p_f^N}{m_N^*}
\right| \right) \right] \,.
\end{equation}
The total energy density of the system can be written as
\begin{eqnarray}
\varepsilon & = & \frac{1}{2} m_{\omega}^2 \omega_0^2
+ \frac{1}{2} m_{\rho}^2 \rho_0^2 + \frac{1}{2} m_{\sigma}^2 \sigma^2
+ \frac{1}{2} m_{\sigma^*}^2 \sigma^{*2} + \frac{1}{2} m_{\phi}^2 \phi_0^2
+\frac{3}{4}d\omega_0^4+ U(\sigma) \nonumber \\
& & \mbox{} + \sum_b \varepsilon_b + \sum_l \varepsilon_l + \frac{{B}^2}{8 \pi^2} \,,
\end{eqnarray}
where the last term is the contribution from the magnetic
field.  The general expression for the
pressure is given by  
\begin{eqnarray}
P= \sum_i \mu_i n_i - \varepsilon.
\end{eqnarray}

At the outermost surface of the star, that is at lower densities, the matter is composed of 
only neutrons, protons and electrons. Hence, at the low density regime, only the electrons and 
protons are affected by the magnetic field. Electron being highly relativistic, the number 
of occupied Landau levels by electrons is very large. The field strength under consideration is 
larger than the critical field strength of electron by several orders but very less than the 
critical field strength of protons. Therefore, the number of occupied Landau levels by protons
is large. As the magnetic field increases with the increase of density, the number of occupied
Landau levels gradually decreases for every species.

\section*{Magnetic field in quark phase}

Considering the simple MIT bag model for the quark matter in presence of magnetic field 
we assume that the quarks are non-interacting. The
current masses of u and d quarks are extremely small, e.g.,
$5$ and $10$ MeV respectively, whereas, for s-quark the current quark mass is taken to be
$150$ MeV, unless otherwise stated.

For the same constant magnetic field configuration along the z-axis, 
the single energy eigenvalue is given by\cite{lan}
\begin{equation}
{E_{i}}=\sqrt{{p_i}^2+{m_i}^{2}+2\nu |q_i|B}
\end{equation}
Then the thermodynamic potential in presence of strong magnetic field $B(>B^{(c)}$, 
critical value discussed later) is given by \cite{chakrabarty-sahu}
\begin{equation}
\Omega_i=-\frac{g_i|q_i|BT}{4\pi^2}\int dE_i\sum_{\nu}\frac{dp_i}
{dE_i}\ln [1+exp(\mu_i-E_i)/T].
\end{equation}
For the zero temperature, the Fermi distribution is approximated by a step function
and by interchanging the order of the summation over $\nu$ and integration
over $E$,
\begin{eqnarray}
\Omega_i&=&-\frac{2g_i|q_i|B}{4\pi^2}\sum_{\nu}\int_{\sqrt{m_i^2+2\nu |q_i|
B}}^{\mu}
dE_i\sqrt{E_i^2-m_i^2-2\nu |q_i|B}.
\label {eq:om}
\end{eqnarray}
The upper limit of $\nu$ sum can be obtained from the following relation
\begin{equation}
{p_{f,i}}^{2}={\mu_i}^2-{m_i}^2-2\nu |q_i|B \ge 0,
\end{equation}
where $p_{f,i}$ is the Fermi momentum of the particle $i$.

The upper limit is not necessarily same for all the components. For a certain critical
magnetic field strength the energy of a charged particle changes significantly in the 
quantum limit. For an electron with mass 0.5 MeV, the critical field strength is 
$\sim 4.4\times 10^{13}$G, whereas for a light quark (u or d), this 
value becomes $\sim 4.4\times 10^{15}$G, and for s-quark of current mass 150 MeV, it is $\sim 10^{19}$G. 
A compact star becomes unstable if the magnetic field strength becomes much greater than $\sim 10^{18}$G,
and so many authors have neglected quantum mechanical effect of magnetic field on s-quarks \cite{cha} but 
in our calculation we include the quantum mechanical effect for all particles.

Assuming the strange quark matter also to be charge neutral and in chemical equilibrium, we may 
write as
\begin{equation}
\mu_d=\mu_s=\mu=\mu_u+\mu_e, \label{eq:ch}
\end{equation}
\begin{equation}
2n_u-n_d-n_s-3n_e=0. \label{eq:cha}
\end{equation}
The baryon number density is given by
\begin{equation}
n_b=\frac{1}{3}(n_u+n_d+n_s) \label{eq:ba}.
\end{equation}
Solving the above eqs(\ref{eq:ch}, \ref{eq:cha}, \ref{eq:ba})
numerically, we obtain the chemical potentials of all the flavors and
electron.
Zero temperature approximation gives the number density of the species $i$
($u,~d,~s,~e$)
\begin{equation}
n_i=\frac{g_i|q_i|B}{4\pi^2}\sum_{\nu}\sqrt{{\mu_i}^2-{m_i}^2-2\nu |q_i|B}.
\end{equation}

The total energy density and pressure of the strange quark matter is given by
\begin{eqnarray}
\varepsilon &=& \sum_{i}\Omega_i +B_G +\sum_{i}n_i \mu_i \nonumber \\
p&=&-\sum_i\Omega_i-B_G,
\end{eqnarray}
where $B_G$ is the bag constant.

\section*{Phase transition and mixed phase}
With the above given hadronic and quark EOS, 
we now perform the Glendenning construction \cite{glen} for the mixed phase,
which determines the range of baryon density where both
phases coexist. 
Allowing both the hadron and quark phases to be 
separately charged, and still preserving the total charge neutrality as a whole in the mixed phase. 
Thus the matter can be treated as a two-component
system, and can be parametrized by two chemical potentials,
usually the pair ($\mu_e, \mu_n$), {\rm i.e.}, electron and 
baryon chemical potential. 
To maintain mechanical equilibrium, the pressure of the two phases are equal.
Satisfying the chemical and beta equilibrium the chemical potential of different
species are connected to each other.
The Gibbs condition for mechanical and chemical equilibrium at zero temperature between both 
phases is given by
\begin{equation}
P_{\rm {HP}}(\mu_e, \mu_n) =P_{\rm{QP}}(\mu_e, \mu_n) = P_{\rm {MP}}. 
\label{e:mp}
\end{equation}
This equation gives the equilibrium chemical potentials 
of the mixed phase corresponding to the intersection of the two phases. 
At lower densities below the mixed phase, the system is in the 
charge neutral hadronic phase, and for higher densities above the mixed phase  
the system is in the charge neutral quark phase. 
As the two surfaces intersect, one can calculate the charge densities $\rho_c^{\rm{HP}}$ and
$\rho_c^{\rm{QP}}$ separately in the mixed phase. If $\chi$ is the volume fraction occupied 
by quark matter in the mixed phase, we have
\begin{equation}
\chi \rho_c^{\rm{QP}} + (1 - \chi) \rho_c^{\rm{HP}} = 0.
\label{e:chi}
\end{equation}

Therefore the energy density $\epsilon_{\rm{MP}}$ and the baryon density 
$n_{\rm{MP}}$ of the mixed phase can be obtained as
\begin{eqnarray}
\epsilon_{\rm{MP}} &=& \chi \epsilon_{\rm{QP}} + (1 - \chi) 
\epsilon_{\rm{HP}}, \\
n_{\rm{MP}} &=& \chi n_{\rm{QP}} + (1 - \chi) 
n_{\rm{HP}}. \label{e:mp1}
\end{eqnarray}

\section*{Results}

\begin{figure}
\vskip 0.2in
\centering
\includegraphics[width=3.0in]{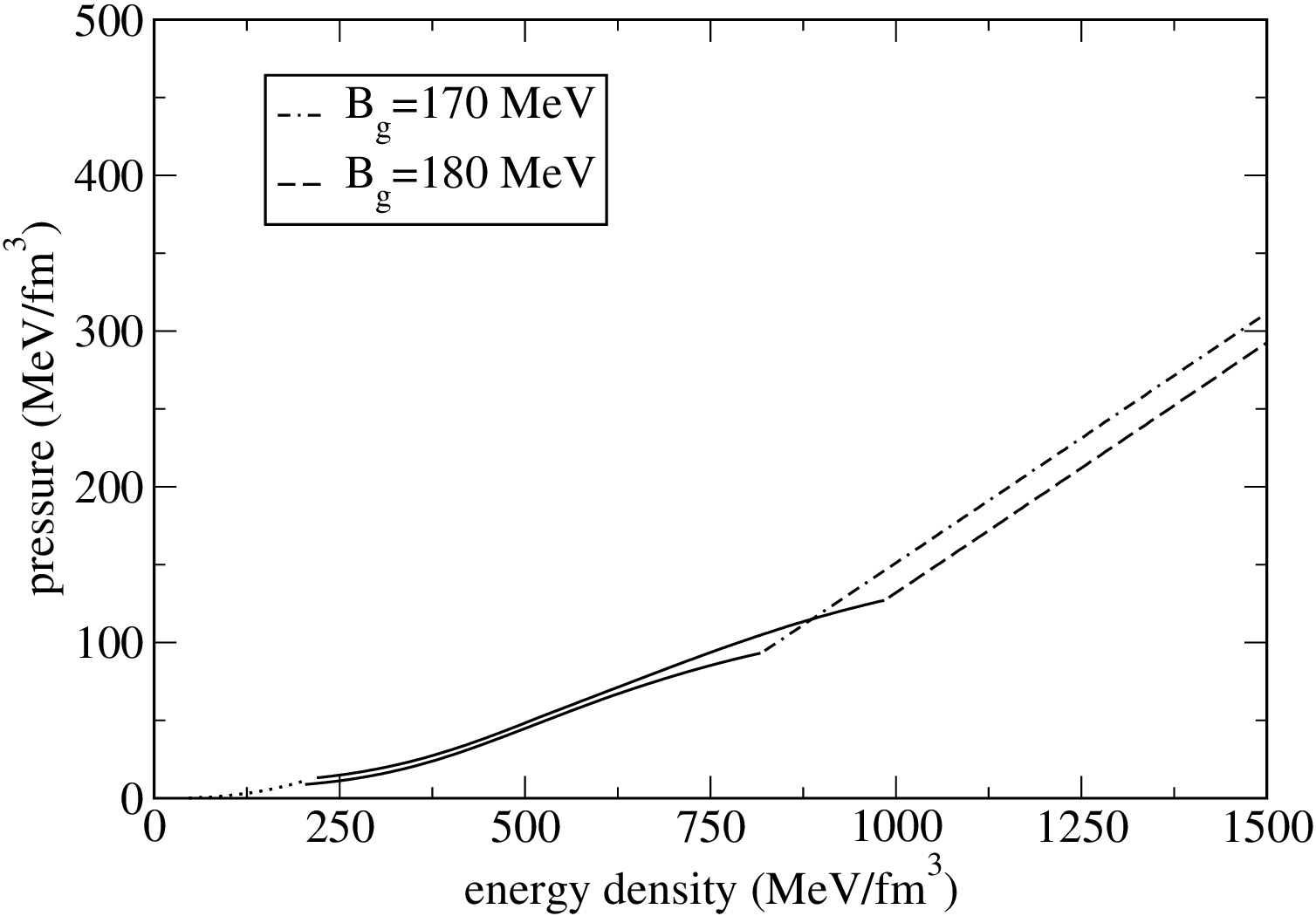}
\caption{Pressure vs energy density plot with bag pressure of $170$ and $180$MeV.}
\label{fig1}
\end{figure}

\begin{figure}
\vskip 0.2in
\centering
\includegraphics[width=3.0in]{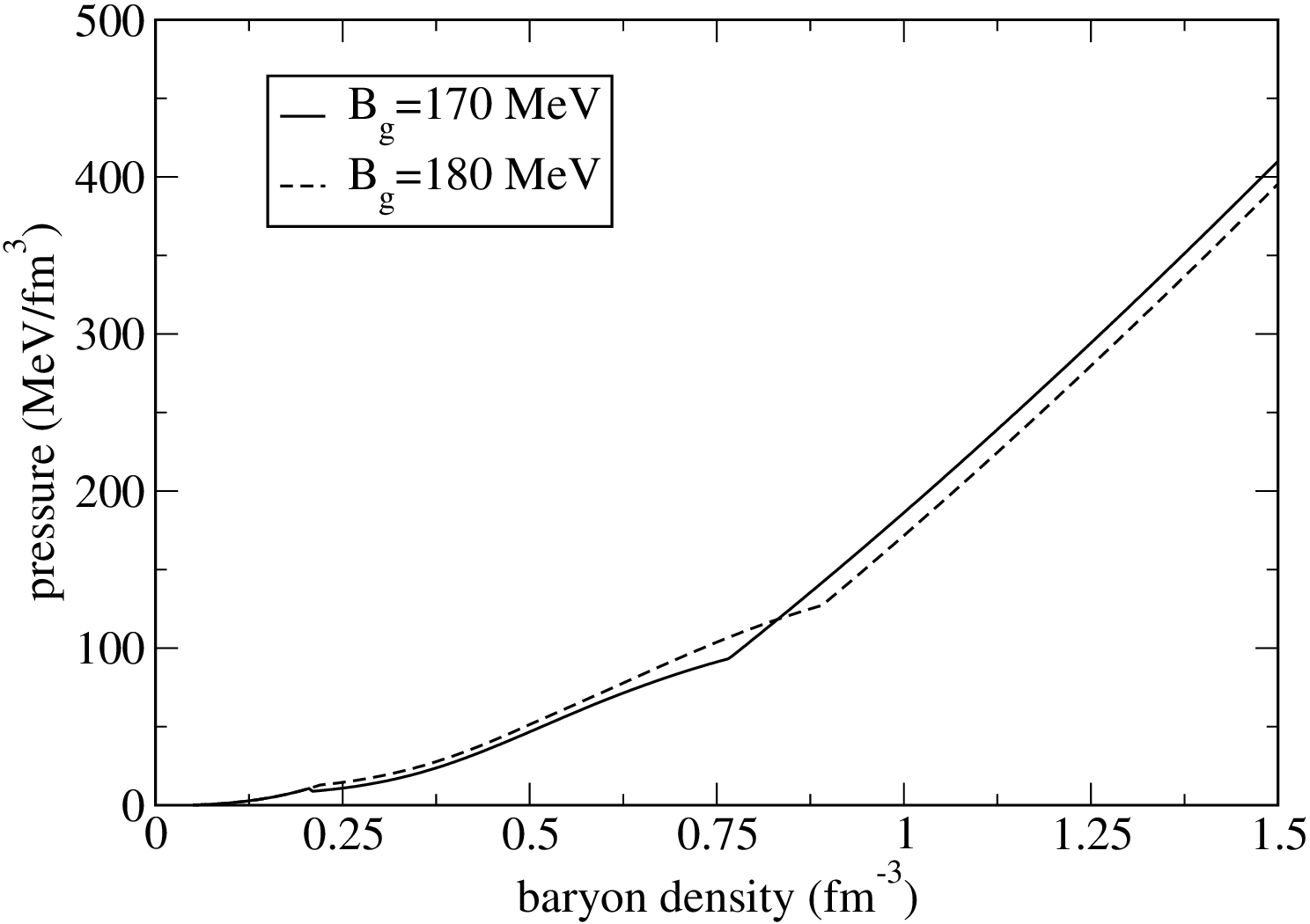}
\caption{Pressure vs number density plot with bag pressure of $170$ and $180$MeV.}
\label{fig2}
\end{figure}

\begin{figure}
\vskip 0.2in
\centering
\includegraphics[width=3.0in]{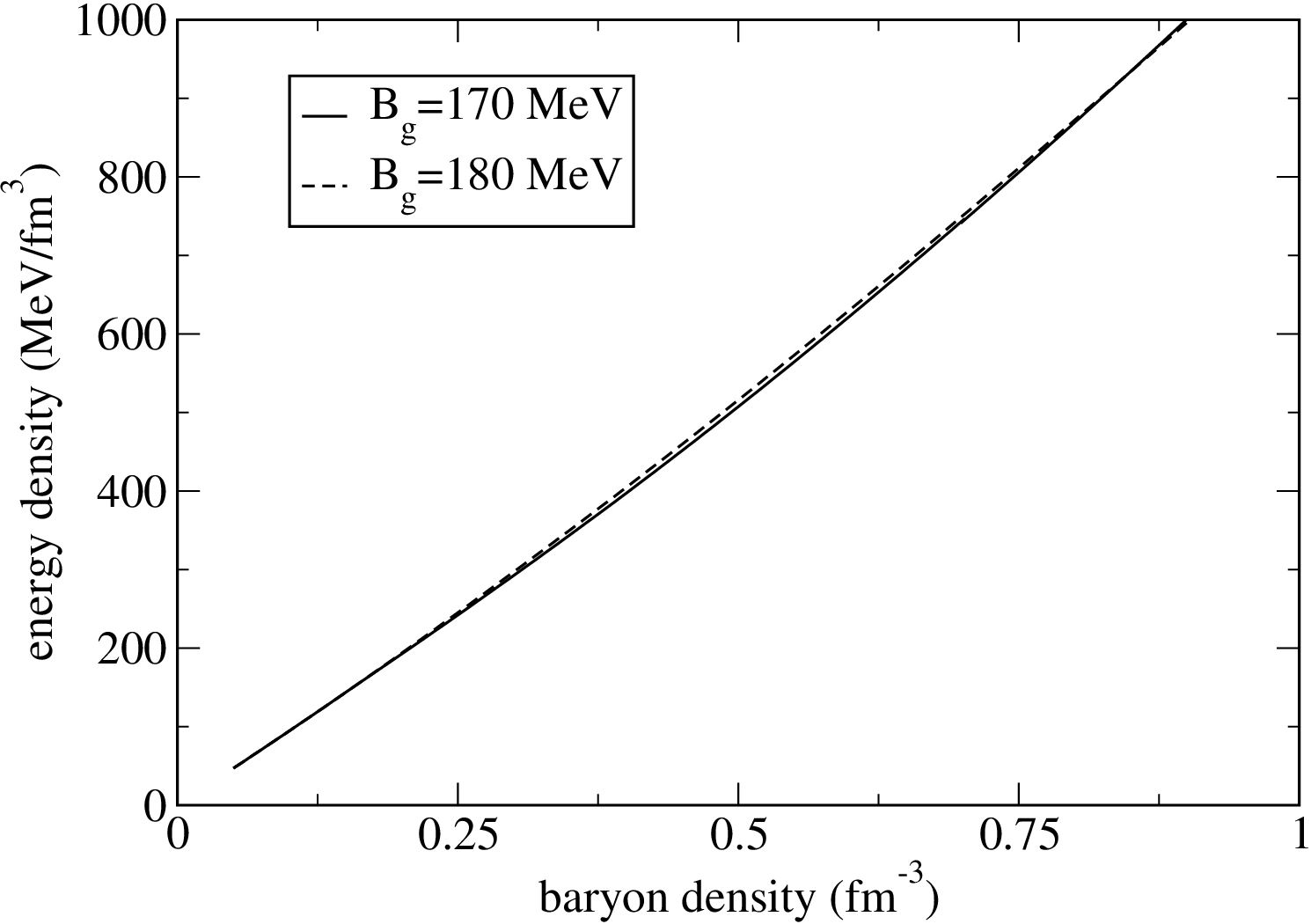}
\caption{Energy density vs number density plot with bag pressure of $170$ and $180$MeV.}
\label{fig3}
\end{figure}

In neutron stars, the central part of the star has maximum 
density, therefore, it is much likely that the matter there undergoes a 
phase transition. As the density decreases towards the surface there is a probability of having
nuclear matter and so in the intermediate stage there is a mixed 
phase, and as we go outwards we only have nuclear matter. The crust consisting 
mainly free electrons and nuclei which completes the star structure.

The parametrization of the EOS of the hadron and quark phase is responsible
for characterization of the mixed phase region. 
For the hadronic EOS we assume a fixed parameter set TM1, which reproduces the nuclear matter
properties at high density quite well. 
However the quark EOS can be controlled by changing the quark masses and the bag constant.
The masses of the light quarks (u and d)
are bounded and we take them to be $5$ and $10$MeV, respectively. The mass of s-quark is still not 
established, and can vary between $100-300$MeV, and we take them to be $150$MeV, unless otherwise stated.
We regulate the bag constant ($B_G$) to characterize the mixed phase region. 
Now we use the Glendenning approach to construct the mixed phase, 
and obtain pressure vs energy density relation as given in fig \ref{fig1}.
In fig \ref{fig1},
we have plotted the mixed phase EOS (pressure vs energy) with bag pressure $170$ and $180$MeV.
For simplicity, we will denote ${B_G}^{1/4}=170 MeV=B_g$.
The lower portion of the curve is nuclear phase (dotted line), the 
intermediate portion is the mixed phase (bold line) and the upper region is the 
quark phase (broken line).
The curve with bag constant $170$MeV is much stiffer than the curve with bag pressure $180$MeV, 
because the bag pressure is negative to the matter pressure, making the effective pressure less. 
In fig \ref{fig2} we have plotted pressure vs number density, and we find that the 
qualitative variation in the curves is same as that of fig \ref{fig1}. 
For bag constant $170$MeV the mixed phase region starts at density $0.2 fm^{-3}$ and ends at
$0.76 fm^{-3}$. With bag constant $180$MeV the mixed phase region starts at density $0.22 fm^{-3}$ and 
ends at $0.89 fm^{-3}$. 
In fig \ref{fig3}
we have plotted for the energy density vs number density, and we find a smooth curve,
which does not differ from each other much. It is clear from the above figures that 
the main variation is due to the pressure, therefore, we only plot the pressure vs
energy density curve. The above curves shows that as the bag pressure 
increases the range of mixed phase region increases, and there is a slight kink in the curve
from going to the quark phase from the mixed phase. The EOS (or the pressure) for the 
nuclear matter is usually much stiffer than quark matter. As the bag constant with $170$ MeV is 
more stiffer than $180$MeV the kink in the former is much sharper than the latter one.
Also as the latter curve is much flatter and so the mixed phase region is much extended there.
By the Glendenning construction, we find that for a given mixed phase to exist the bag constant must be
in between $170$MeV and $180$ MeV.

\begin{figure}
\vskip 0.2in
\centering
\includegraphics[width=3.0in]{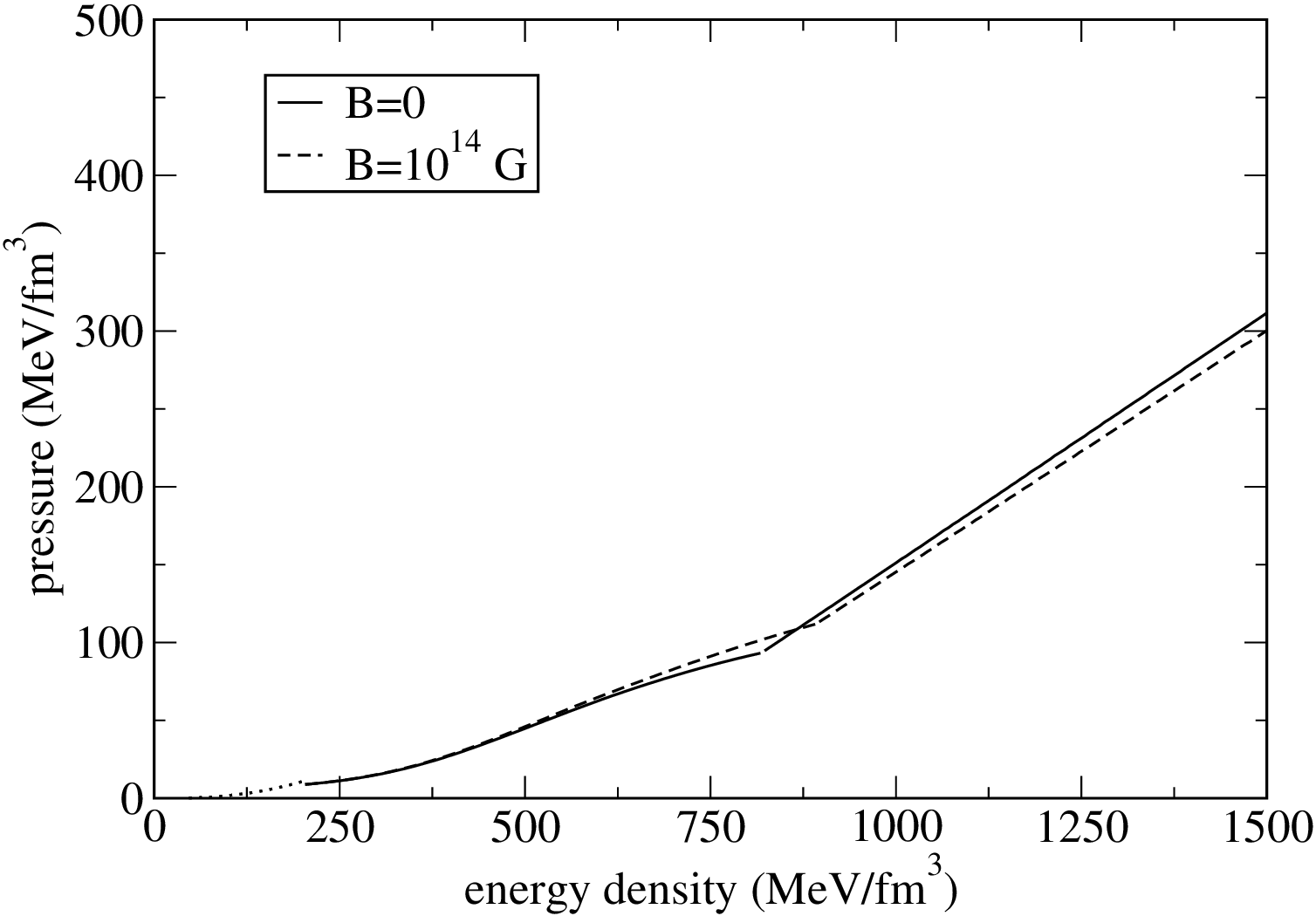}
\caption{Pressure against energy density plot with bag pressure of $170$MeV with and without magnetic field.
The magnetic field is $B=10^{14}$G.}
\label{fig4}
\end{figure}

\begin{figure}
\vskip 0.2in
\centering
\includegraphics[width=3.0in]{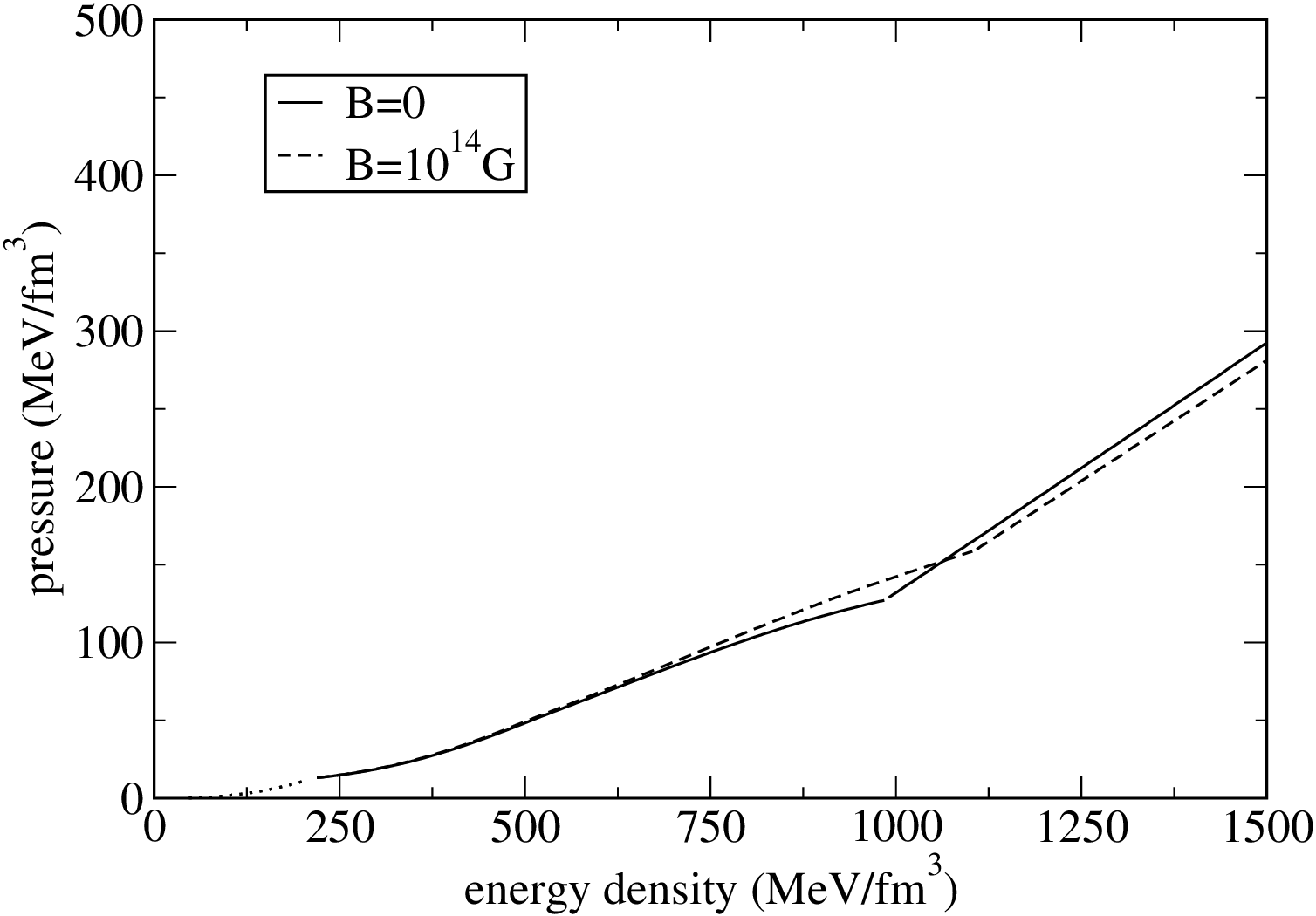}
\caption{Pressure against energy density plot with bag pressure of $180$MeV with and without magnetic field.
The magnetic field is $B=10^{14}$G.}
\label{fig5}
\end{figure}

\begin{figure}
\vskip 0.2in
\centering
\includegraphics[width=3.0in]{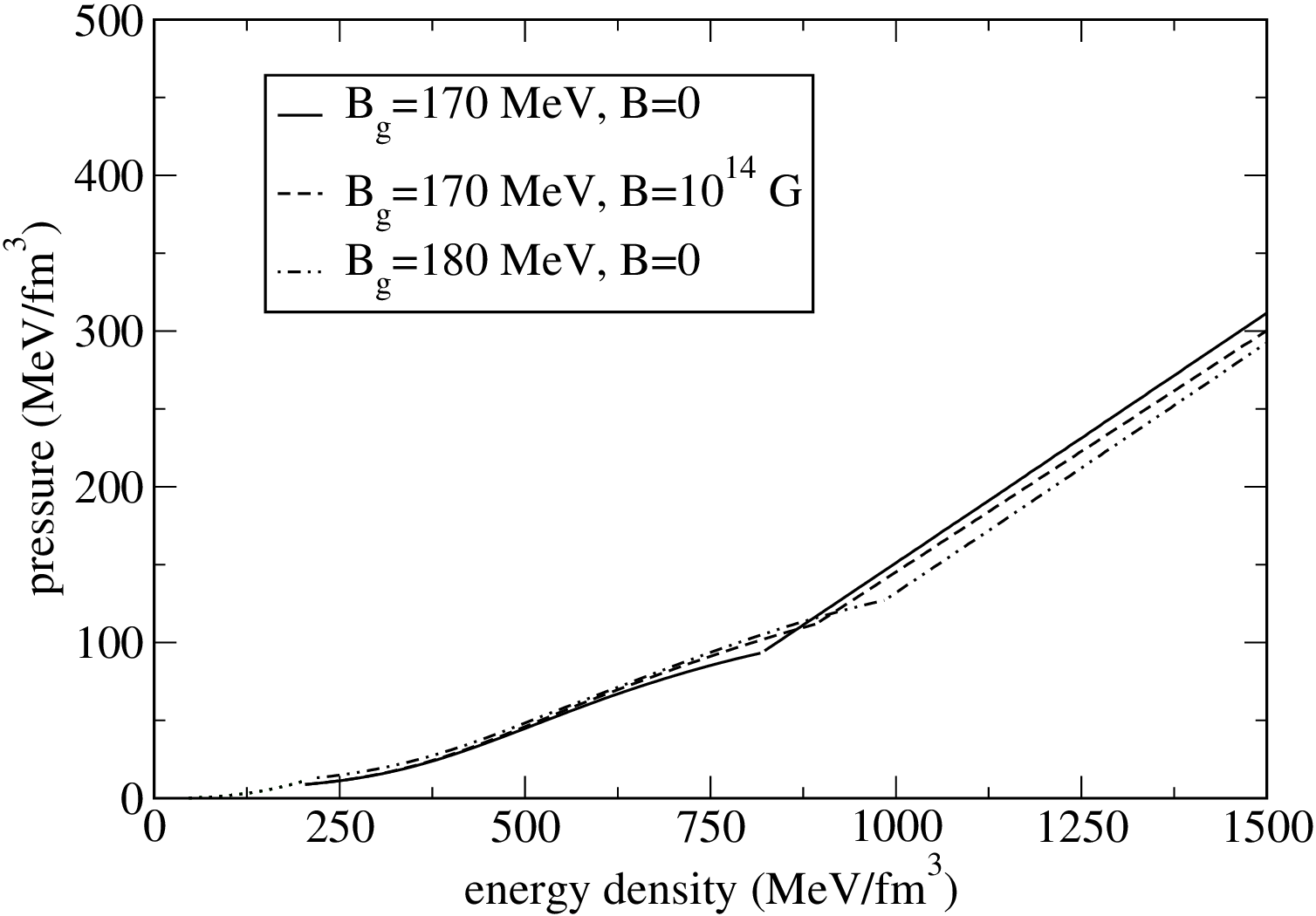}
\caption{Pressure against energy density plot with bag pressure of $170$ and $180$MeV. For the bag pressure 
$170$MeV we have also plotted curve with magnetic field of strength $10^{14}$G.}
\label{fig6}
\end{figure}

The introduction of the magnetic field changes the EOS of the matter.
The single particle energy is now Landau quantized, and thereby it changes all 
the other thermodynamic variable of the EOS, namely the number density, 
pressure and the energy density. In fig \ref{fig4}, we have plotted 
EOS for bag constant $170$MeV with and without the magnetic field. The effect of 
magnetic field is insignificant when the field strength is less than $10^{14}$G, and also 
for this case the effect in the nuclear phase is very small. The magnetic field effect is less for the 
nuclear matter than quark matter because, the nuclear EOS is much steeper than 
the quark EOS, thereby requiring much greater field to have any sound effect.
We have plotted the same for bag constant $180$MeV (fig \ref{fig5}), and for comparing the two
bag constants, we have plotted fig \ref{fig6}. For the bag constant $170$ MeV, the EOS curve with 
magnetic field extends up to density $0.8 fm^{-3}$, and for $180$ MeV it extends upto density
$0.92 fm^{-3}$. The change in the mixed phase region is about $5-7 \%$.
Magnetic field makes the curve softer due to the negative effect of 
landau quantization on the matter pressure
and the positive effect on the matter energy density. As shown in the figures, the effect 
of magnetic field begins to appear on the EOS of the matter when the field strength is above
$10^{14}$G. Such field has very less effect on the nuclear matter but has considerable
effect on the mixed and quark matter. With the onset of the magnetic field the mixed phase region gets 
extended. The magnetic field makes the quark matter EOS more flatter and therefore 
the mixed phase region is much extended. For fixed bag constant and fixed magnetic field 
value throughout, we cannot go to field strength above $10^{15}$G, as it is bounded by observation
of surface magnetic fields in magnetars.

Next we assume a density dependent bag constant. In the literature
there are several attempts to understand the density dependence of bag constant \cite{adami,blaschke}; 
but still there is no definite picture, and most of them are model dependent. 
We parametrized the bag constant in such a way that it attains a value $B_\infty$, 
asymptotically at very high densities. The experimental range of $B_{\infty}$ is given
in Burgio et al. \cite{burgio1,burgio}, and from there we choose the value $B_{\infty}=130$MeV. 
With such assumptions we then construct a Gaussian parametrization given as \cite{burgio1,burgio}
\begin{eqnarray}
B_{gn}(n_b)  =  B_\infty  +  (B_g  -  B_\infty)  \exp  \left[  -\beta  \Big(
\frac{n_b}{n_0} \Big)^2 \right] \:. \label{e:g}
\end{eqnarray}
The lowest value $B_{\infty}$, is the lowest value of bag pressure which it attains at asymptotic 
high density in quark matter, and is 
fixed at $130$MeV. The quoted value of bag pressure, is the value of the bag constant 
at the nuclear and mixed phase intersection point denoted by $B_g$ in the equation. 
The value of $B_{gn}$ decrease with
increase in density and attain $B_\infty=130$MeV asymptotically, the rate of decrease 
of the bag pressure is governed by parameter $\beta$.

The observed magnetic field of the magnetars is of the order of $\sim 10^{14}-10^{15}$G.
The flux conservation of the progenitor star may give the central field as high as 
$\sim 10^{17}-10^{18}$G.We assume that the parametrization of the magnetic field 
depends on the baryon number density.
Therefore we assume a simple density dependence, given by \cite{chakrabarty97,monika}
\begin{equation}
{B}(n_b)={B}_s+B_0\left\{1-e^{-\alpha \left(
\frac {n_b}{n_0} \right)^\gamma}\right\},
\label{mag-vary}
\end{equation}
where $\alpha$ and $\gamma$ determines the magnetic field varaiation for fixed
surface field $B_s$ and asymptotic central field $B_0$. The value of $B$ depends mainly on $B_0$,
and is quite independent of $B_s$. Therefore we vary $B_0$, whereas surface 
field strength is taken to be fixed at $B_s=10^{14}$G. We keep $\gamma$ fixed at $2$, and 
vary $\alpha$ for to have the field variation. Previous authors considered very high magnetic
field value at the center, few times $10^{18}$G, but we would assume the maximum field to be
of the order of few $10^{17}$G. As this is somewhat low value from other previous assumptions,
but it is more likely to be present in most magnetars.

\begin{figure}
\vskip 0.2in
\centering
\includegraphics[width=3.0in]{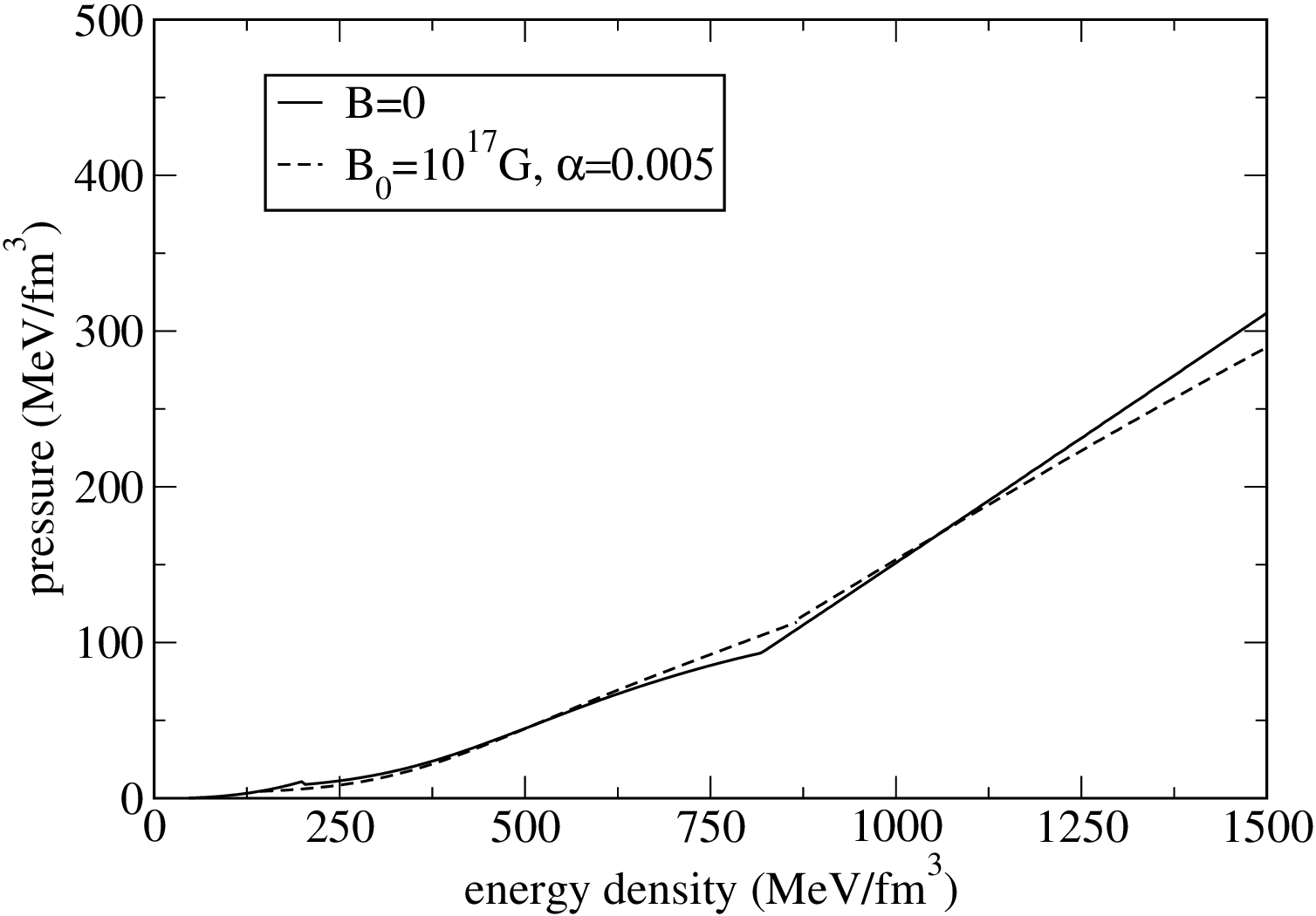}
\caption{Pressure as energy density plot with bag pressure of $170$MeV, with and without varying 
magnetic field. The varying magnetic field has $B_0=10^{17}$G and $\alpha=0.005$.}
\label{fig7}
\end{figure}

Next, we vary the  magnetic field, and the bag constant is kept constant.
First we consider the bag constant to be $B_g=170$MeV. In fig \ref{fig7}
we have plotted curves for zero magnetic field and with $B_0=10^{17} G$
with $\alpha=0.005$. As we vary the magnetic field, the magnetic 
field increases as we go towards to the center of the star. The field quoted in the figure 
is asymptotic field value. With $B_0=10^{17}$G and $\alpha=0.005$, the field strength
is $4 \times 10^{16}$G at $10 n_0$.
It is clear from the figure as the field strength increases, the curve becomes less stiffer. 
The change in the curve stiffness is 
due to the fact that the magnetic pressure due to landau quantization act in the 
opposite direction of the matter pressure, whereas, for the magnetic stress it acts
towards the matter energy density. The two effect reduces the stiffness of the EOS 
(pressure vs energy density curve).
It is also clear that the nuclear region (the low density regime) is not much affected by the 
magnetic field as there the magnetic field strength is low, whereas
the quark sector (higher density regime) is the most effected region as the field strength is maximum there. 
However, the mixed phase region is moderately affected (the intermediate region). 

\begin{figure}
\vskip 0.2in
\centering
\includegraphics[width=3.0in]{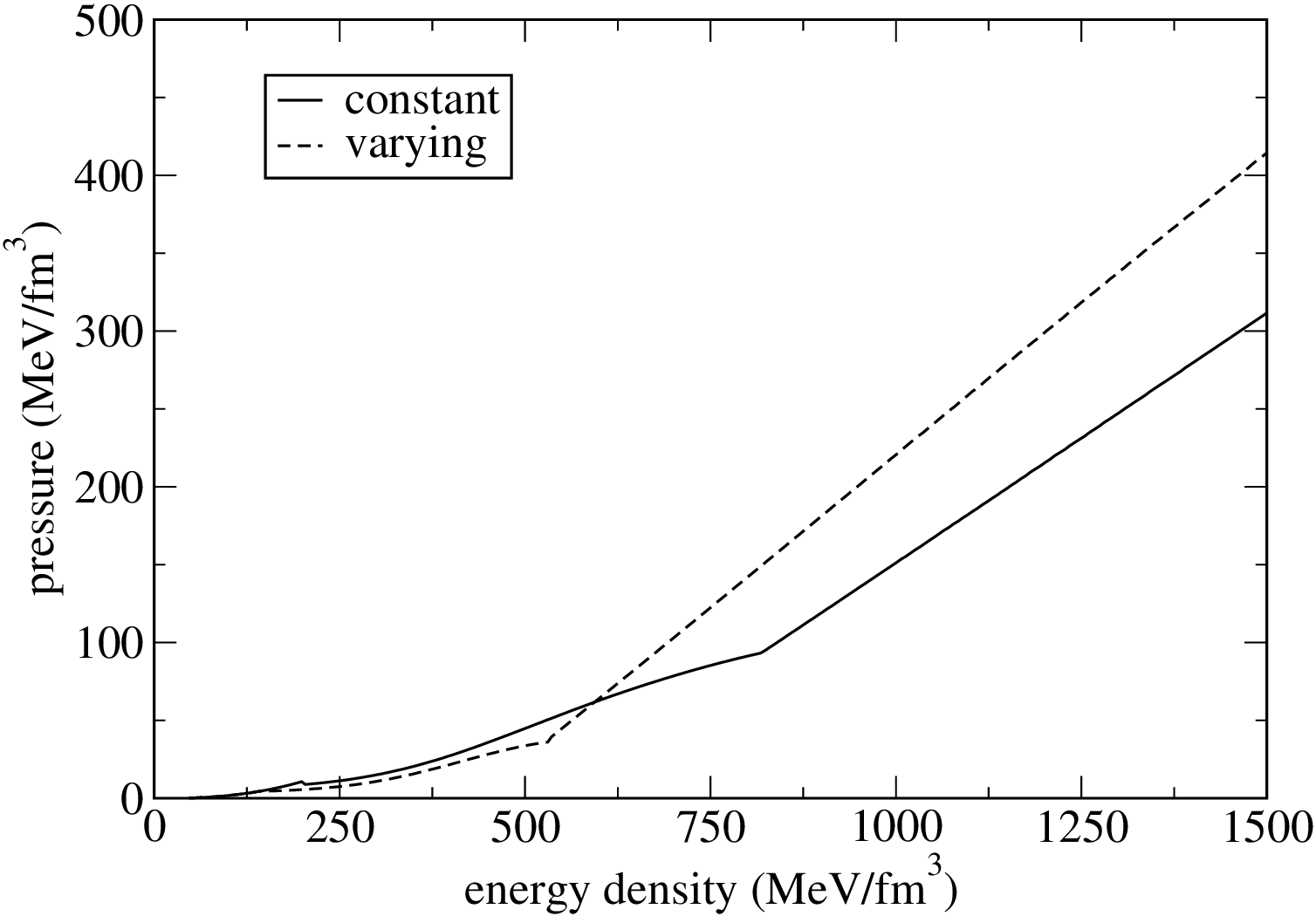}
\caption{Pressure against energy density plot with constant and varying bag pressure, $B_g=170$MeV.}
\label{fig8a}
\end{figure}

\begin{figure}
\vskip 0.2in
\centering
\includegraphics[width=3.0in]{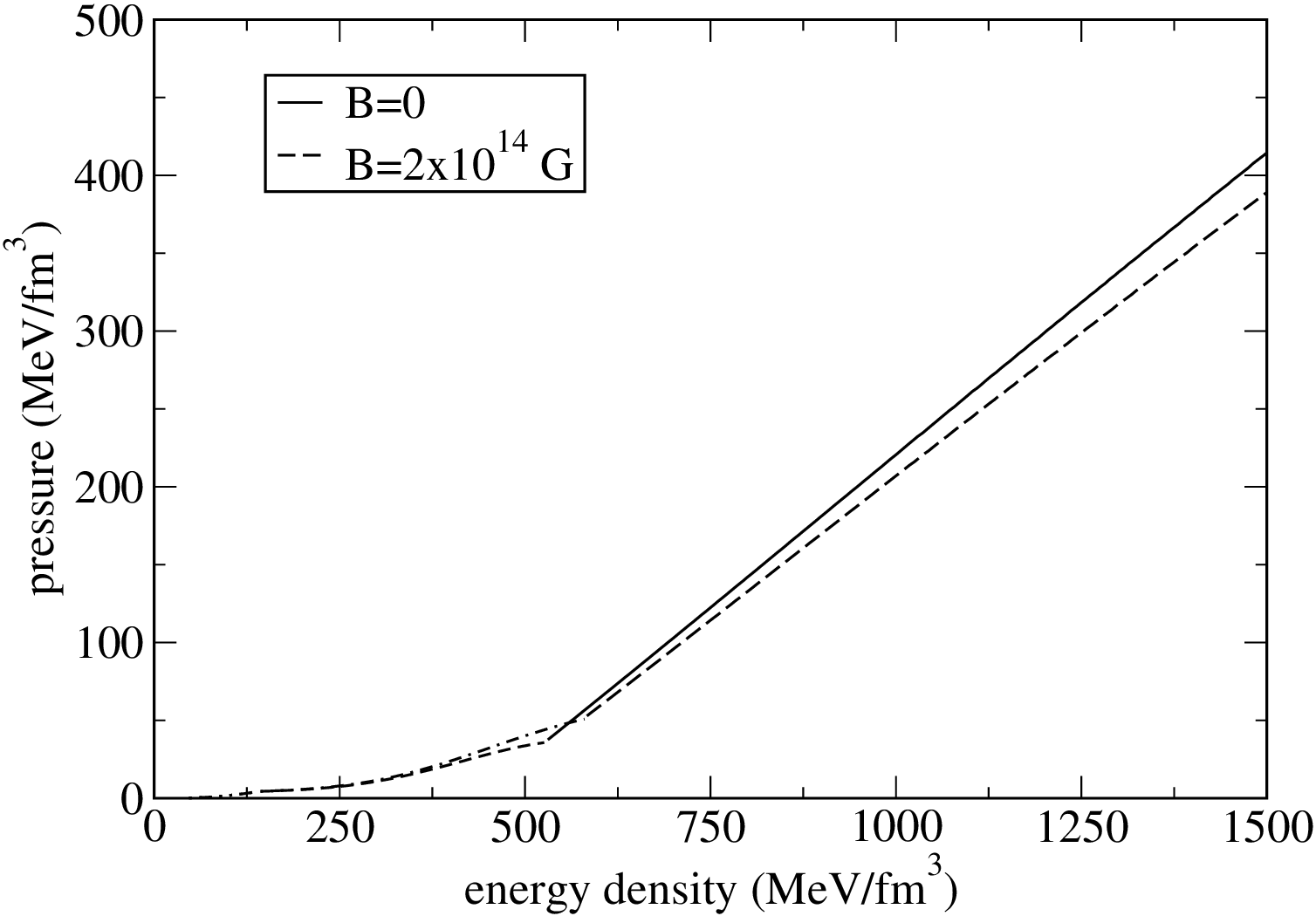}
\caption{Pressure against energy density plot having density dependent bag pressure $B_g=170$MeV, 
with and without magnetic field. The magnetic field strength is $B=2\times 10^{14}$G.}
\label{fig9}
\end{figure}

\begin{figure}
\vskip 0.2in
\centering
\includegraphics[width=3.0in]{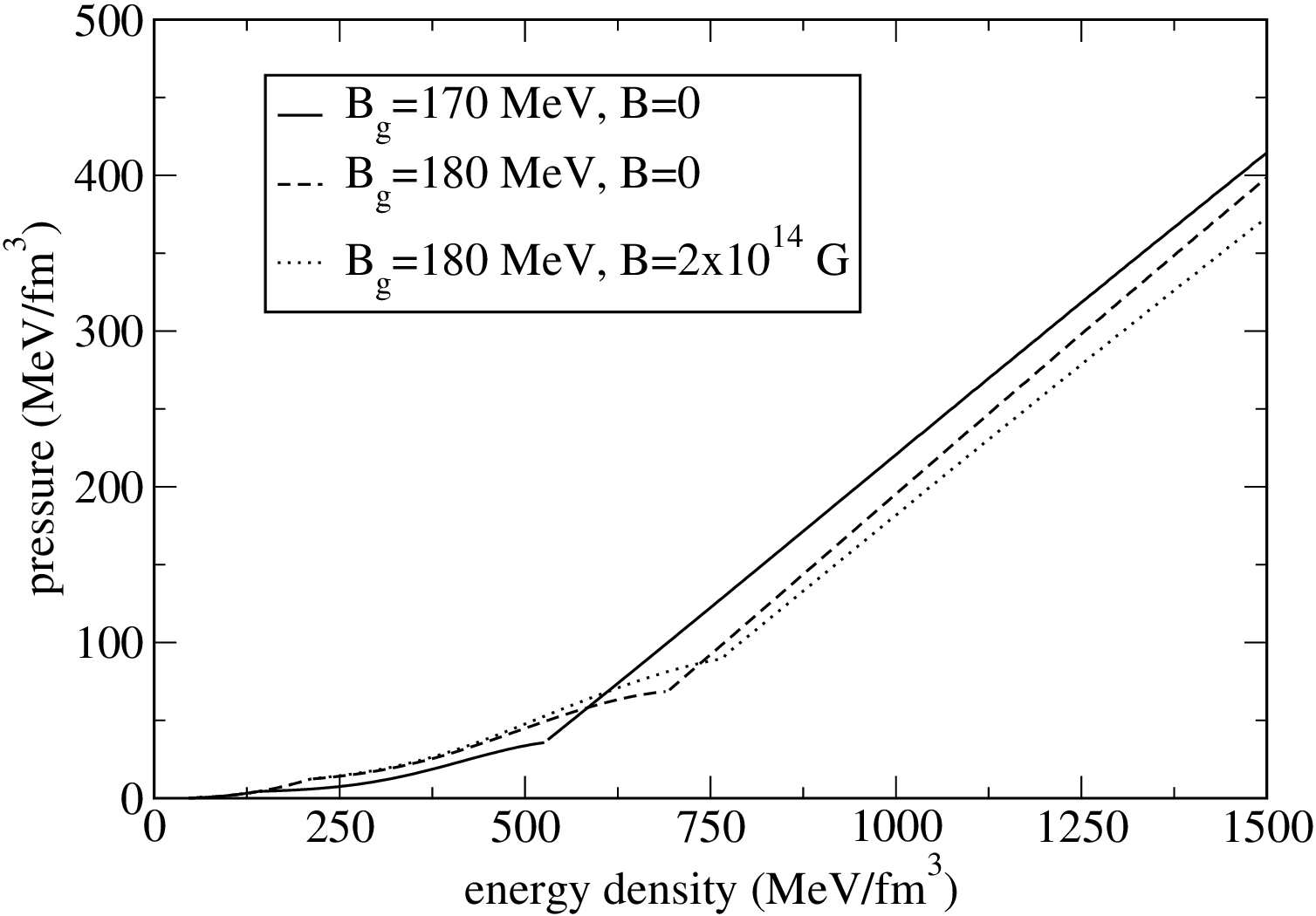}
\caption{Pressure against energy density plot having density dependent bag pressure $B_g=170$MeV and 
$180$MeV. Also shown in the figure the magnetic field ($B=2\times10^{14}$G) induced EOS curve for $B_g=170$MeV.}
\label{fig10}
\end{figure}

In fig \ref{fig8a} we plot curves with and without varying bag pressure, $B_g=170$MeV. 
For the curve with variation, at higher densities the bag pressure decreases, making the effective 
matter pressure higher. Therefore the pressure against energy density plot for this case is much stiffer.
Also the mixed phase region gets shrunken due to the varying bag pressure. The mixed phase region now
only extends up to density $0.53 fm^{-3}$. The change in the mixed phase region is about $40 \%$.
Therefore the change in the mixed phase region is much more influenced by varying bag pressure
than due to magnetic field.
We have plotted fig \ref{fig9} with varying bag pressure $B_g=170$MeV, with and without constant magnetic field.
The magnetic field employed for this plot is $2\times10^{14}$G. 
The change in the slope of the curves is due to the Landau quantization effect. The magnetic pressure
acts opposite to the matter pressure, making the curve flat.
For comparison, we have plotted fig \ref{fig10} with density dependent bag pressure, 
$B_g=170$ and $B_g=180$MeV, and obtain quantitative same result.  

\begin{figure}
\vskip 0.2in
\centering
\includegraphics[width=3.0in]{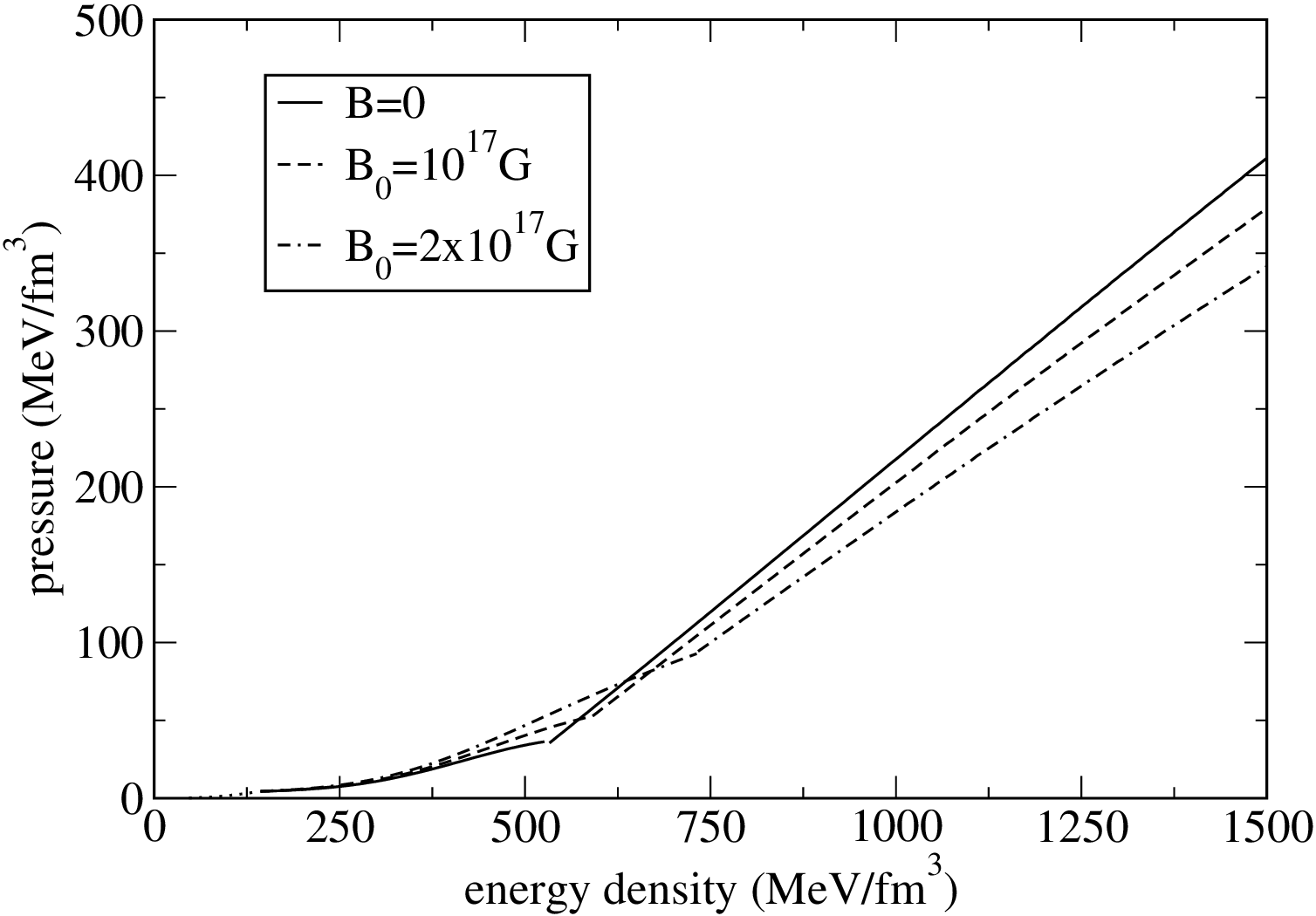}
\caption{Pressure with energy density plot having density dependent bag pressure $170$MeV, without magnetic 
field and with two different magnetic fields, having $\alpha=0.005$.}
\label{fig11}
\end{figure}

\begin{figure}
\vskip 0.2in
\centering
\includegraphics[width=3.0in]{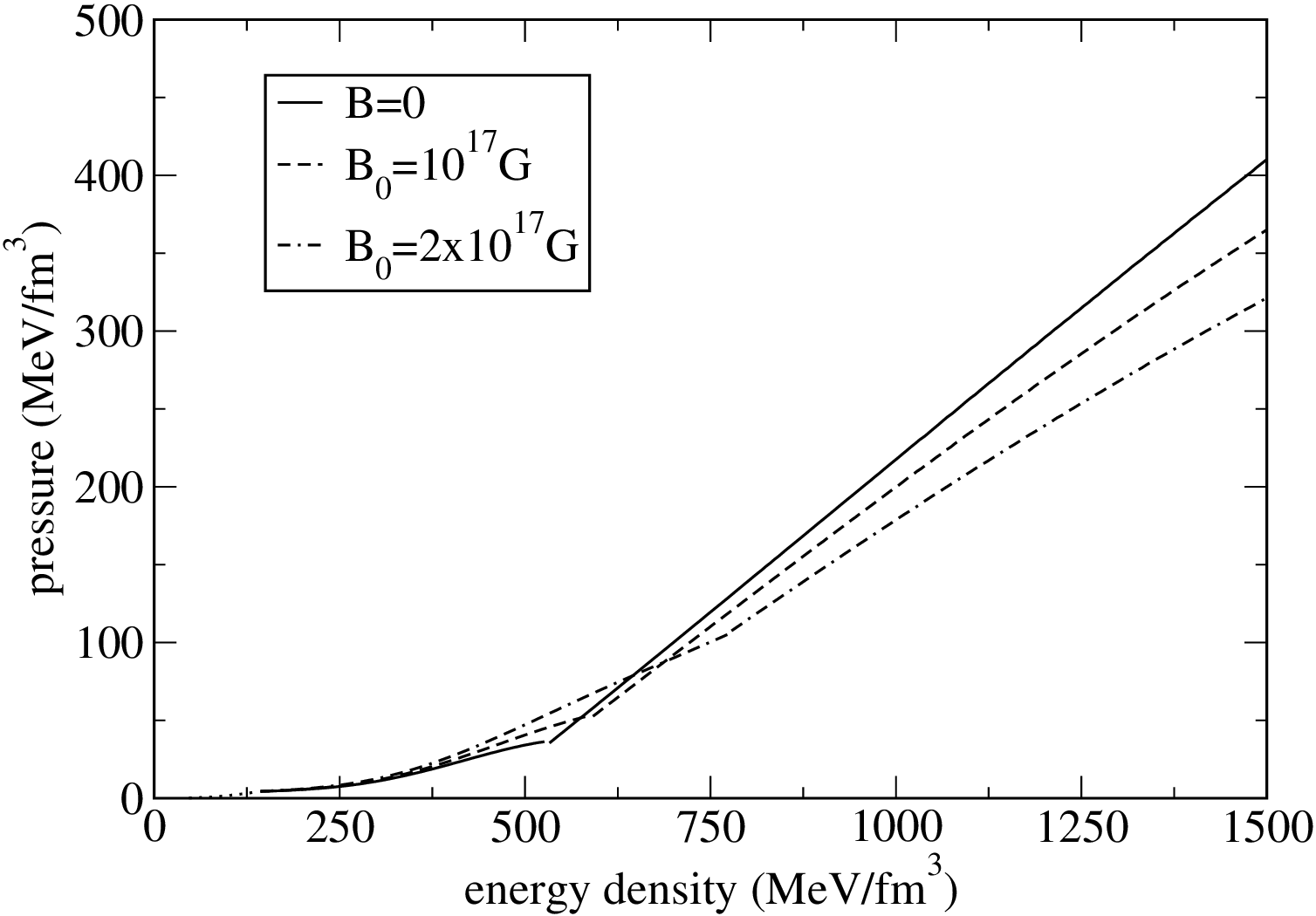}
\caption{Pressure with energy density plot having density dependent bag pressure $170$MeV,  without magnetic 
field and with two different magnetic fields, having $\alpha=0.01$.}
\label{fig12}
\end{figure}

\begin{figure}
\vskip 0.2in
\centering
\includegraphics[width=3.0in]{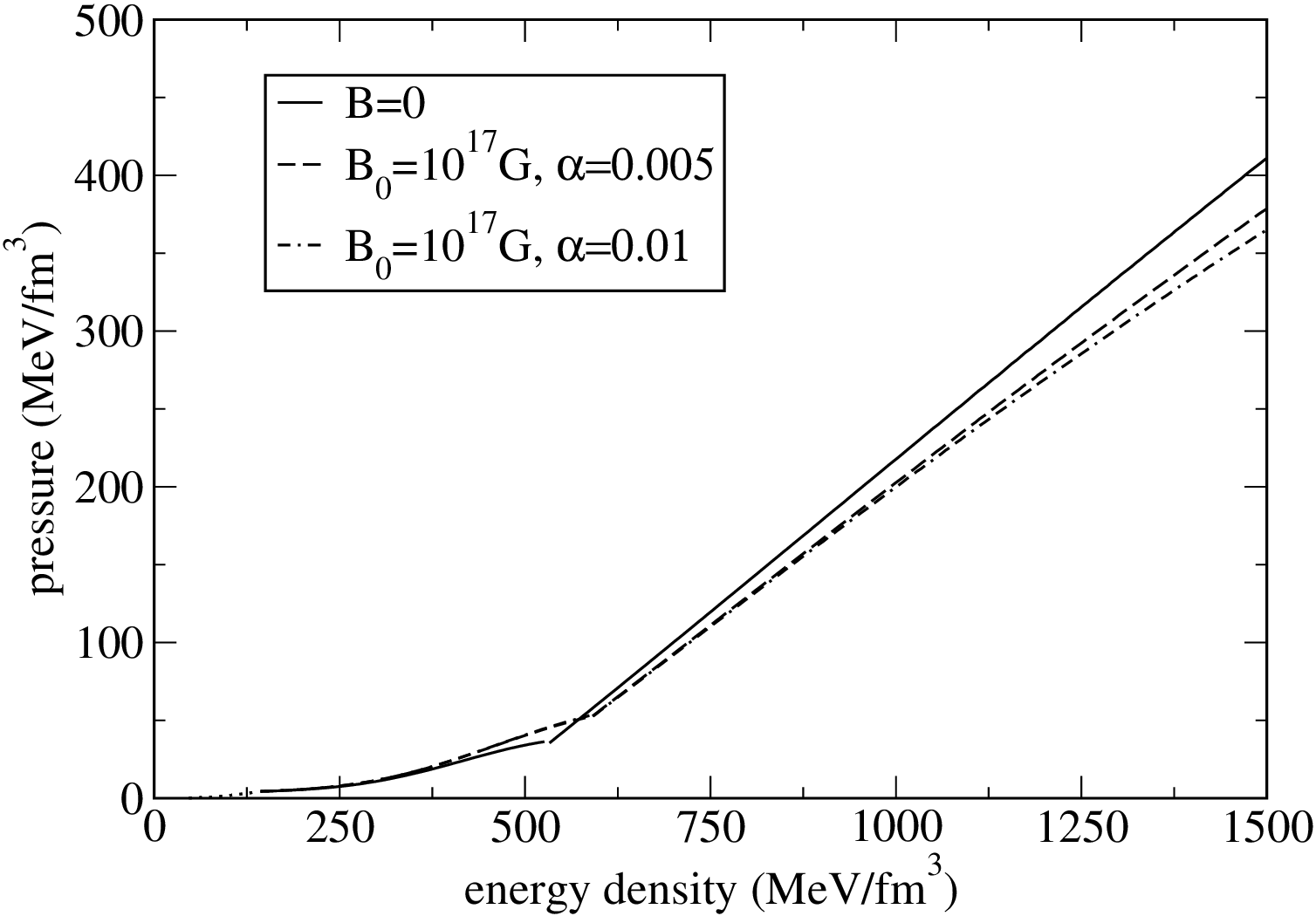}
\caption{Pressure with energy density plot having density dependent bag pressure $170$MeV, without magnetic 
field and with same magnetic field but different $\alpha$ values.}
\label{fig13}
\end{figure}

The curves for which both the bag constant and the magnetic field varies are of
utmost importance.  
Fig \ref{fig11} shows curves for varying bag pressure $170$MeV, without magnetic field
and with varying magnetic field, $B_0=10^{17}$G and $2 \times 10^{17}$G having $\alpha=0.005$.
For the above values the field strength is $4 \times 10^{16}$G and
$7.8 \times 10^{16}$G, at density $10n_0$. As the value of $B_0$ increases the slope of the EOS curves
becomes more and more soft, because the value of magnetic pressure increases with increase
in field strength. As the magnetic pressure increases the effective pressure decreases making 
the curves flatter. 
In fig \ref{fig12} we plot the same set of curve only for $\alpha=0.01$. 
With such $\alpha$ value, the asymptotic $B_0=10^{17}$G gives field strength of $6 \times 10^{16}$G 
at $10n_0$ baryon density.  For $B_0=2 \times 10^{17}$G the field strength is $1.21 \times 10^{17}$G 
at the same $10n_0$ baryon density.
As the variation ($\alpha$) becomes stiffer, the EOS curve becomes softer. This is 
seen clearly in fig \ref{fig13}. 

\begin{figure}
\vskip 0.2in
\centering
\includegraphics[width=3.0in]{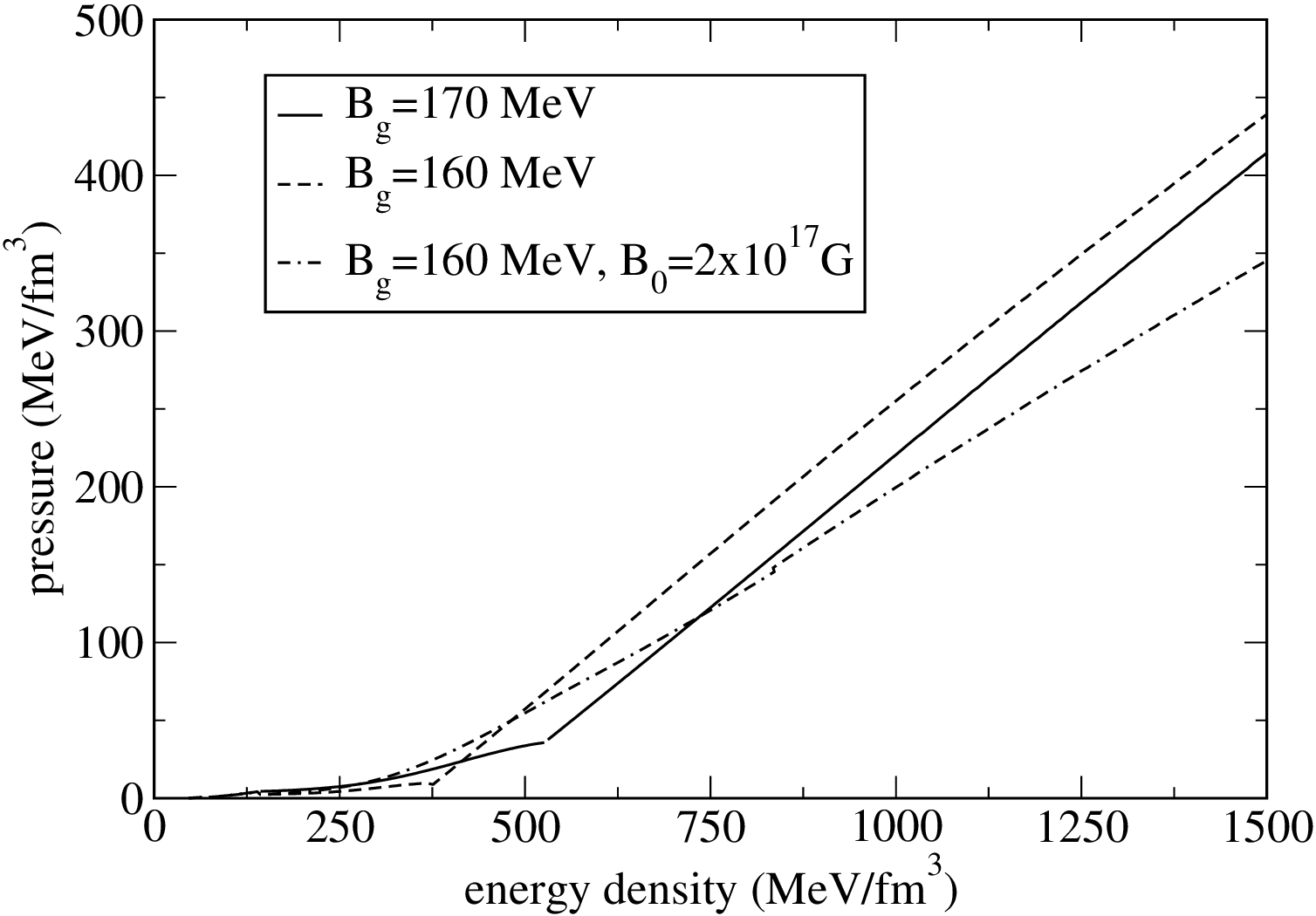}
\caption{Pressure vs energy density plot with two different density dependent bag pressure 
$160$MeV, $170$MeV. We have also plotted the magnetic field induced (field strength $B_0=2\times 10^{17}$G) 
EOS curve for bag pressure $160$MeV having $\alpha=0.01$.}
\label{fig15a}
\end{figure}
We find for such varying bag constant and varying magnetic field, the change in the curves
from the non varying non magnetic case is maximum. There is considerable change
in the stiffness of the curves and also change in the mixed phase region.  
Towards the center, the magnetic field increases whereas the bag pressure decreases. On one hand the 
low bag pressure makes the curve stiffer whereas on the other hand large magnetic field strength
makes the curve flatter. The low bag constant makes the mixed phase region to shrink, and the larger
magnetic field tries to expand the mixed phase region. The effect of bag pressure is greater than the 
magnetic field and therefore the mixed phase is smaller than the constant bag pressure case. 
On the low density side, the
effect of magnetic field is insignificant. Therefore the phase boundary
between the nuclear and mixed phase is not much affected. 

For a varying bag constant we can have a significant mixed phase region with $B_g=160$MeV 
(fig \ref{fig15a}). The curve with bag pressure $160$MeV is stiffer than other curves.
This is because the bag pressure of $B_g=160$MeV is lower than other higher bag pressure.
Therefore, the effective matter pressure for this curve is higher than any other curve, which
is reflected in the stiffness of the curve. For bag constant $160$MeV the mixed phase region 
starts at density $0.15 fm^{-3}$ and ends at $0.38 fm^{-3}$. 

Assuming the star is non rotating and has spherically symmetric, the distribution of mass is in
hydrostatic equilibrium. The equilibrium configurations solution are obtained
by solving the Tolman-Oppenheimer-Volkoff (TOV) equations \cite{shapiro} for 
the pressure $P(\epsilon)$ and the enclosed mass $m$,
\begin{widetext}
\begin{eqnarray}
  {dP(r)\over{dr}} &=& -{ G m(r) \epsilon(r) \over r^2 } \,
  {  \left[ 1 + {P(r) / \epsilon(r)} \right] 
  \left[ 1 + {4\pi r^3 P(r) / m(r)} \right] 
  \over
  1 - {2G m(r)/ r} } \:,
\\
  {dm(r) \over dr} &=& 4 \pi r^2 \epsilon(r) \:,
\end{eqnarray}
\end{widetext}
$G$ is the gravitational constant. 
Starting with a central energy density $\epsilon(r=0) \equiv \epsilon_c$,  
we integrate out until the pressure on the surface equals the one 
corresponding to the density of iron.
This gives the stellar radius $R$ and the total gravitational mass is then 
\begin{equation}
M_G~ \equiv ~ m(R)  = 4\pi \int_0^Rdr~ r^2 \epsilon(r) \:. 
\end{equation}
For the description of the NS crust, we have added the hadronic 
equations of state with the ones by Negele and Vautherin \cite{negele}
in the medium-density regime, and the ones   
by Feynman-Metropolis-Teller \cite{feynman} and Baym-Pethick-Sutherland 
\cite{baym} for the outer crust. 

\begin{figure}
\vskip 0.2in
\centering
\includegraphics[width=3.0in]{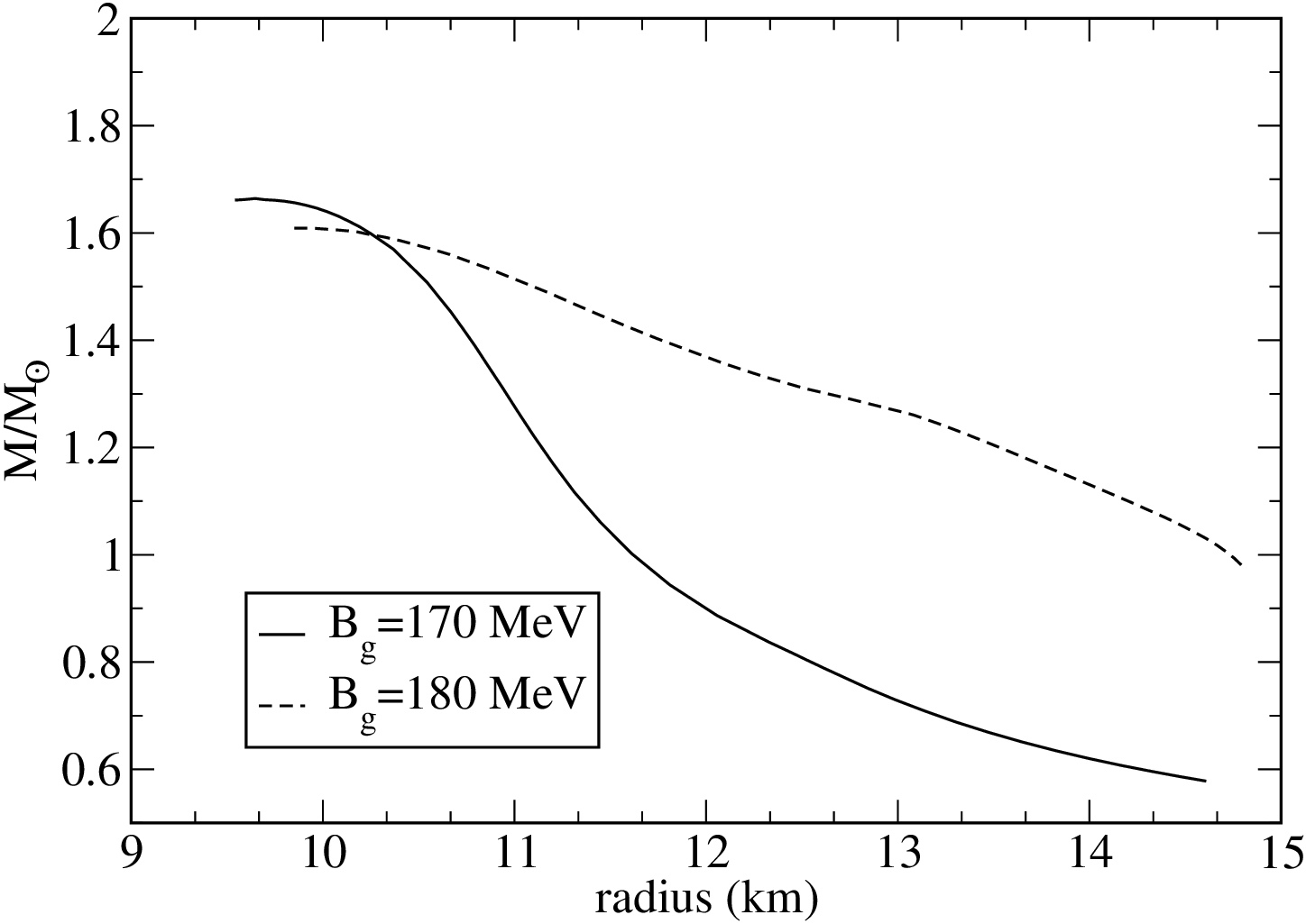}
\caption{Gravitational mass (in solar mass) against radius plot of a star sequence with two different 
density dependent bag pressure, $170$ and $180$MeV.}
\label{fig17}
\end{figure}

\begin{figure}
\vskip 0.2in
\centering
\includegraphics[width=3.0in]{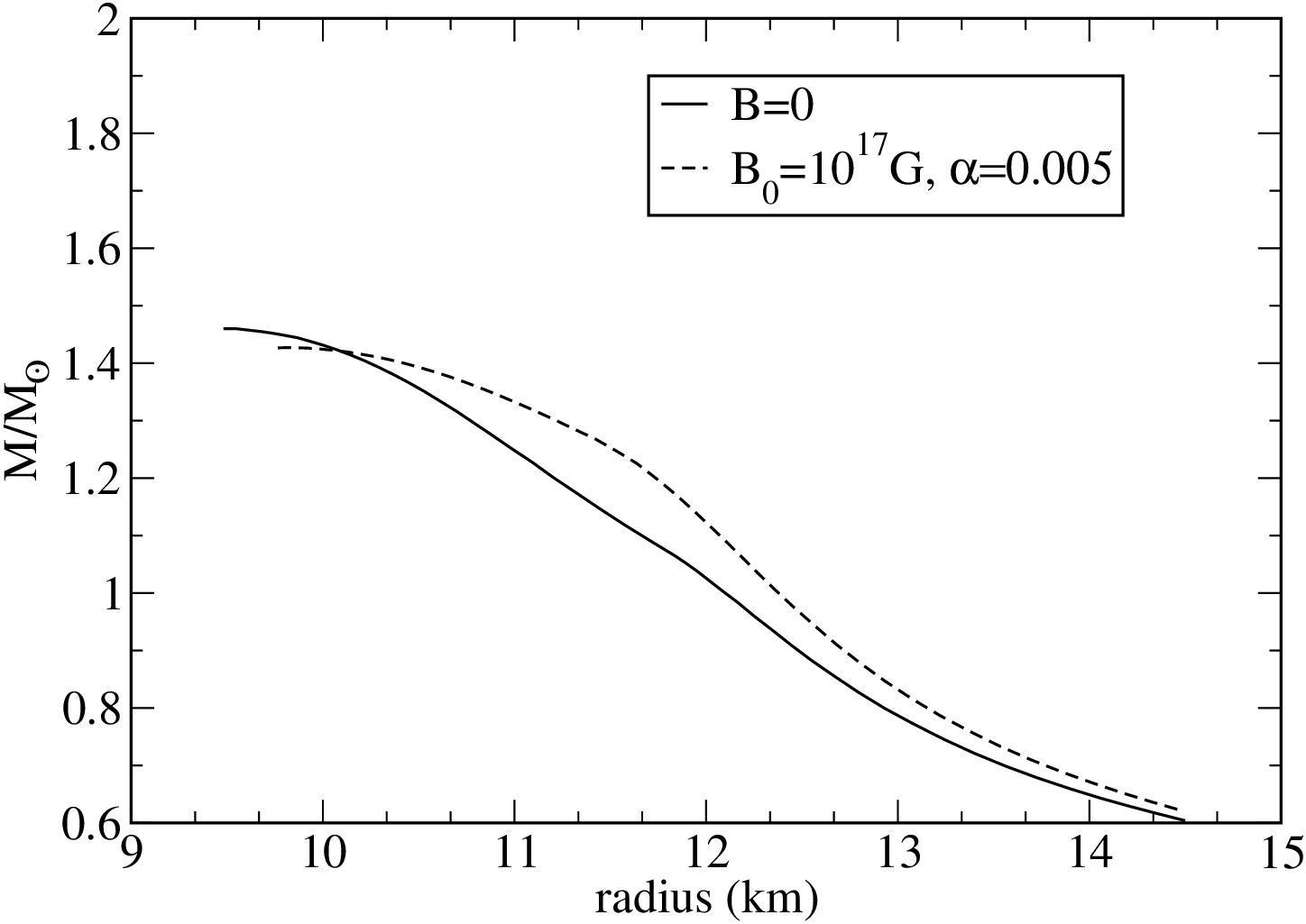}
\caption{Gravitational mass (in solar mass) with radius plot of a star sequence with constant bag pressure 
of $170$MeV, without magnetic field and with varying magnetic field of field strength $B_0=10^{17}$G 
and $\alpha=0.005$.}
\label{fig18}
\end{figure}

Fig \ref{fig17} shows the gravitational mass $M$ (in units of solar mass $M_{\odot}$)
as a function of radius $R$, for varying bag pressure $B_g=170$ and $180$MeV. 
As the bag pressure increases the curve becomes flat as the 
effective matter pressure decreases (bag pressure being negative) thereby decreasing 
the maximum mass of the star. We notice that a flatter EOS corresponds to a flatter mass-radius curve. 
Next we plot fig \ref{fig18}, with constant bag pressure of $170$MeV, with and without magnetic field.
The mass vs radius curve in fig \ref{fig18} is flatter than fig \ref{fig17} because this 
corresponds to the EOS for constant bag pressure, which is much flatter than the EOS with 
varying bag pressure. The varying magnetic field has $B_0=10^{17}$G and $\alpha=0.005$. 
Initially, the mass for the star with magnetic field is higher, but the maximum mass is lower
than the non magnetic case, because the non magnetic EOS is steeper than the magnetic counterpart.
The stiffness (or flatness) of the pressure vs energy density curve for a particular EOS is reflected
in the stiffness (or flatness) of the corresponding mass-radius curve.

\begin{figure}
\vskip 0.2in
\centering
\includegraphics[width=3.0in]{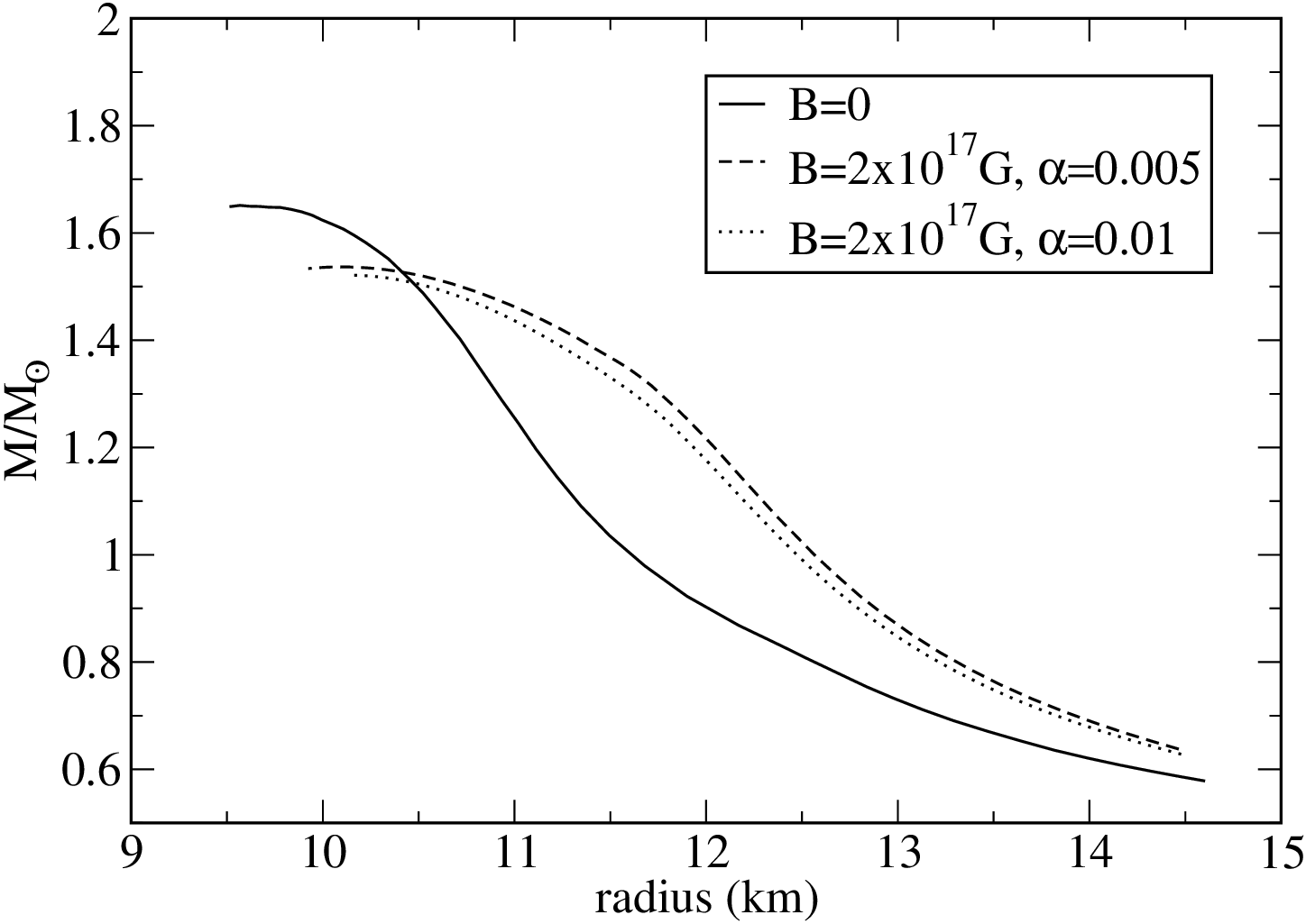}
\caption{Gravitational mass (in solar mass) with radius plot of a star sequence having density dependent bag 
pressure $170$MeV. Curves are plotted without magnetic field and with same magnetic field, of field strength 
$B_0=2\times 10^{17}$G but different $\alpha$.}
\label{fig19}
\end{figure}

\begin{figure}
\vskip 0.2in
\centering
\includegraphics[width=3.0in]{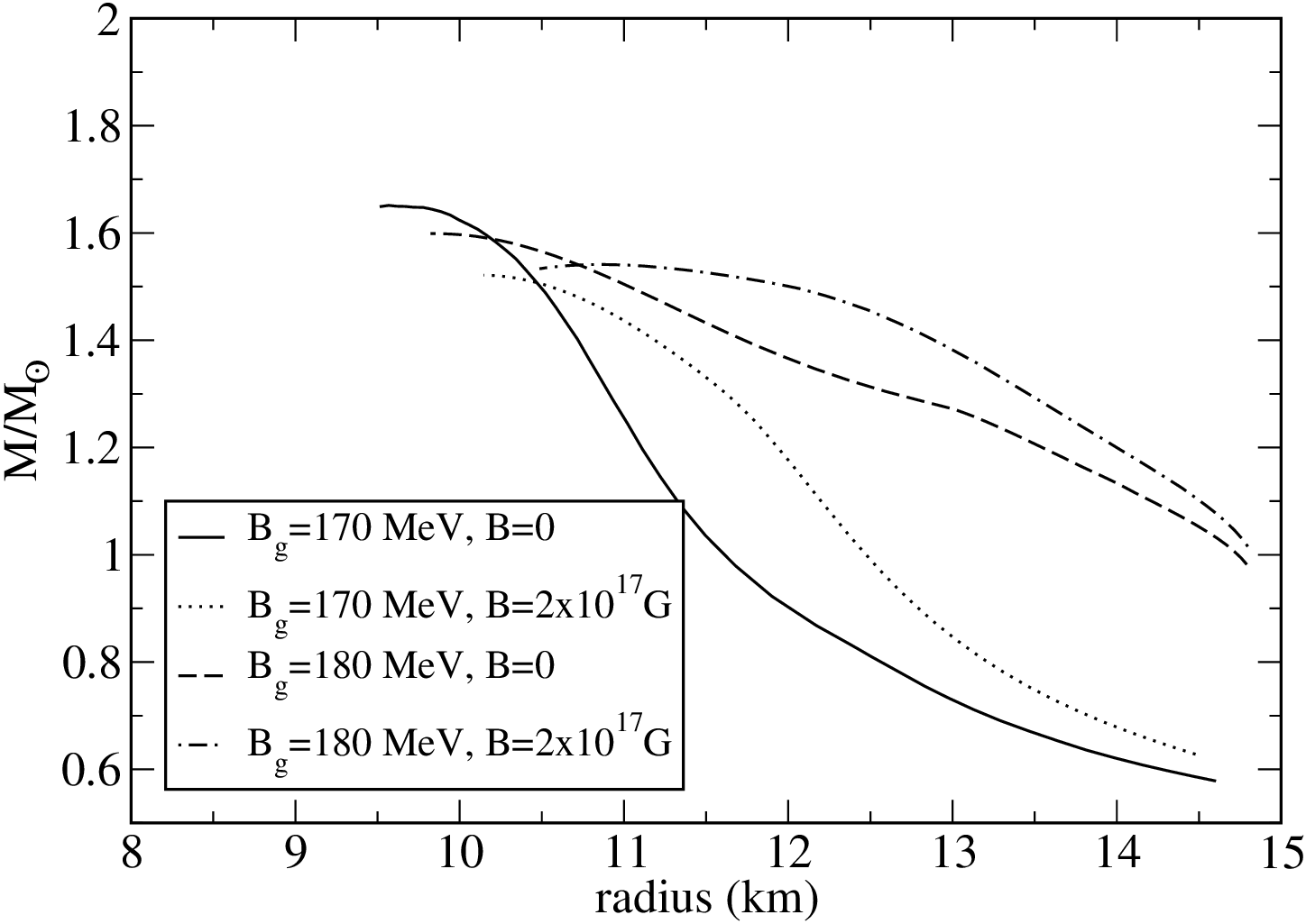}
\caption{Gravitational mass (in solar mass) against radius plot of a star sequence with two different varying 
bag pressure of $170$MeV and $180$MeV. The curves are plotted without magnetic field and with magnetic 
field, of strength $B_0=2\times 10^{17}$G having same $\alpha=0.01$.}
\label{fig20}
\end{figure}

\begin{figure}
\vskip 0.2in
\centering
\includegraphics[width=3.0in]{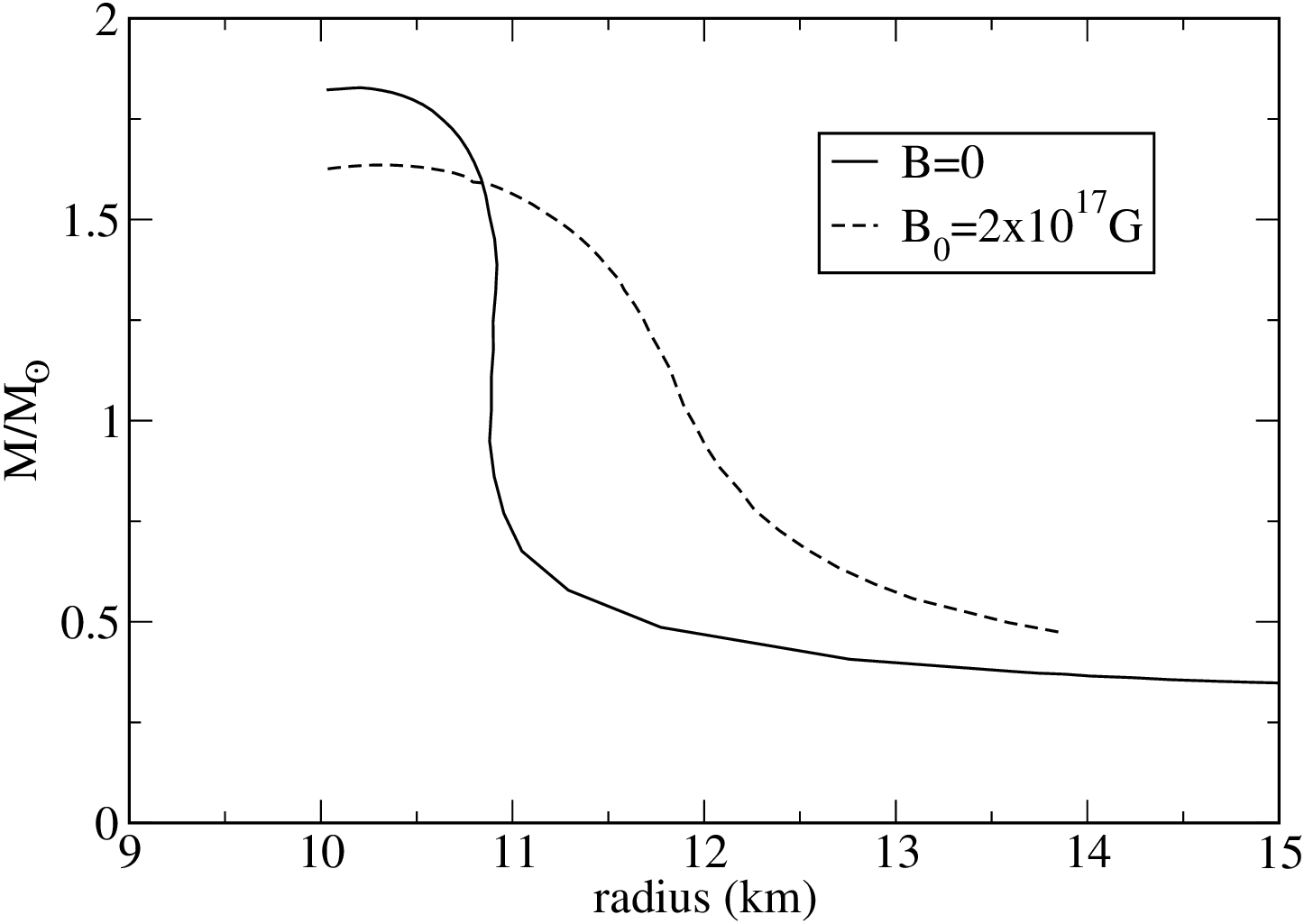}
\caption{Gravitational mass (in solar mass) against radius plot of a star sequence with density dependent 
bag pressure of $160$MeV. Two curves are plotted, one without magnetic field and one with 
magnetic field, having strength $B_0=2\times 10^{17}$G and $\alpha=0.01$.}
\label{fig21}
\end{figure}

Next in fig \ref{fig19}, we plot curves with varying bag constant ($170$MeV) for two different value 
of alpha ($0.005$ and $0.01$), with field strength of $B_0=2\times10^{17}$G. 
Both the magnetic field and bag pressure are density dependent.
The magnetic field makes the mass-radius curve flatter.
As the magnetic field variation becomes
higher, increasing the magnetic field strength as we go inwards, and thereby making the EOS flat. 
As the EOS becomes flat the mass-radius curve also becomes flat, and the maximum mass decreases. 
T compare the mass dependence
on varying magnetic field and varying bag pressure we have plotted curves for two different set of 
curves with varying bag
pressure $170$ and $180$MeV (fig \ref{fig20}). Each set comprising of two curves one without magnetic 
field and one with magnetic field, of strength $B_0=2\times10^{17}$G. 
The qualitative nature of the curves remains same due the reasons discussed earlier. 
As it has been pointed out, with varying bag constant and varying magnetic field we can have 
mixed phase EOS with bag
pressure of $160$MeV. In fig \ref{fig21}, we have plotted the mass-radius curve for $B_g=160$MeV, with 
($B_0=2\times10^{17}$G) and without magnetic field. The magnetic field is varying having $\alpha$ of $0.01$.
The maximum mass for this case is obtained without the 
magnetic field effect and the introduction of the magnetic field makes the curve flatter and also 
reduces the maximum mass. The maximum mass
of a mixed hybrid star obtained with such mixed phase region is $1.84 M_{\odot}$.     

Recently, after the discovery of high-mass pulsar PSR J1614-2230 \cite{Demorest10} with mass of about
$1.97 M_{\odot}$, the EOSs describing the interior of a compact star have been put to severe
constraint. The \cite{Demorest10} typical values of the central density of J1614-2230, for the
allowed EOSs is in the range 2$n_0$ - 5$n_0$. On the other hand, consideration 
of the EOS independent analysis of \cite{lattimer2005} sets the upper limit of 
central density at $10n_0$. For a constant bag pressure, the mass of the HS is about
$1.5$ solar mass (fig \ref{fig18}). With a varying bag pressure, the maximum mass limit can be
increased. The maximum mass limit of mixed phase EOS star with the above given set of 
EOS, with strange mass of $160$MeV is calculated to be $1.84$ solar mass. The maximum mass 
for the mixed hybrid star can be increased to $2.01$ \cite{ritam12a} solar mass with s-quark mass 
of $300$MeV and varying bag pressure of $150$ MeV. 
Therefore the mass limit set by the observation of pulsar PRS J1614-2230 can be 
maintained by the mixed hybrid star having density dependent bag constant. 
But for this particular choice, the mixed phase region is very small.
It should be mentioned here that this mass
limit is only for this set of nuclear and quark matter EOSs. Using very stiff EOS sets (hadronic NL3
and quark quark NJL model) the maximum mass limit for the mixed hybrid star can be raised much higher
as pointed by Lenzi \& Lugones \cite{lenzi}.   

The main aim of this paper was to show the effect of magnetic field on the mixed phase EOS and its 
effect on the maximum mass of a star. We were also interested in showing whether simple EOS 
(hyperonic nuclear and MIT bag quark) can reach the limit set by PSR J1614-2230. 
The other most interesting fact of this calculation is that the mixed hybrid star has radius 
corresponding to the maximum mass, quite different from
the nuclear and strange star. They are not as compact as strange stars and their 
radius lies between the nuclear and strange star. It is also clear from
our calculation that, if the magnetic field influence the EOS only through the 
Landau quantization, it has a negative effect on the matter pressure thereby making the
EOS softer, and the star becomes less massive.

\section*{Summary and discussion}

To summarize, we have studied the effect of magnetic field on the nuclear and
quark matter EOS. We have taken into account Landau quantization effect on
the charged particles of both the EOS. We have considered relativistic mean field 
EOS model for the nuclear matter EOS. For the quark
matter EOS, we have considered simple MIT bag model with density dependent bag
constant. The nuclear matter EOS is much stiffer than the quark matter EOS, and so  
the effect of magnetic field is much more pronounced in the quark matter. 
The magnetic field due to Landau quantization 
softens the EOS for both the matter phases since the  magnetic pressure contributes 
negatively to the matter pressure. 
Here we should mention that the effect of magnetization of matter is important for 
strong magnetic fields, however it is believed that in NS such magnetization is 
mall \cite {broderick}. Therefore in our calculation we have neglected such effect.

Glendenning construction \cite{glen}, determines the range of baryon density where both
phases coexist. At densities below the mixed phase, the system is in the 
charge neutral hadronic phase, and for densities above the mixed phase the system is in charge neutral
quark phase. We have considered density dependent bag pressure, which has been parametrized according
to the Gaussian form. We have fixed the lowest 
value of the bag pressure to be $130$MeV, known from the experiments \cite{burgio}. Accordingly, we have 
also considered varying magnetic field.
Observationally, the inferred surface magnetic field of a NS may be as 
high as $10^{15}$G and is believed to increase at the center. As the density 
decreases with increasing radial distance, we have taken the parametrization
of the magnetic field as a function of density, having maximum field strength at the core.
Considering density dependent bag pressure and 
magnetic field, we construct mixed phase EOS following Glendenning construction.

We find that the effect of magnetic field is insignificant unless the surface field 
is of the order of $10^{14}$G. Such constant magnetic field value has no effect
on the nuclear matter EOS and has very little effect on the mixed and quark matter EOS.
For a varying magnetic field whose surface value is $10^{14}$G but whose central value is 
of the order of $10^{17}$G, we find significant effect on the stiffness of the EOS and also on the 
the extend of the mixed phase region in the EOS. 
As the bag pressure increases the EOS for the quark phase becomes soft, and hence
more the effect of magnetic field. At the central region, the bag pressure decreases but the magnetic 
field increases, and so their respective effect on the EOS act in the opposite direction. 

The magnetic field increases as we go to much higher densities, and so the boundary between
the mixed phase and the quark phase changes with increasing field strength. 
As the magnetic field increases, the EOS becomes less stiffer
and the phase boundary between the mixed and quark phase shifts upwards to the higher 
density value. Towards the low density regime of the curve the
effect of magnetic field is less pronounced, as the magnetic field strength is less 
and also the nuclear matter EOS is much stiffer. Therefore the phase boundary
between the nuclear and mixed phase is less affected.

The maximum mass limit of mixed phase EOS star is also shown 
in this paper. We obtain a significant mixed phase region    
with central bag constant of $160$MeV having s-quark mass of $150$MeV. For higher s-quark mass
($300$MeV) we get a small mixed phase region with bag pressure $150$MeV. For such
a case we find the maximum mass for a mixed hybrid star with the given set of EOS is $2.01 M_{\odot}$. 
The maximum mass is obtained without magnetic field effect
and the introduction of the magnetic field reduces the maximum mass.
Therefore the mass limit set by the observation of pulsar PRS J1614-2230 is 
maintained by the mixed star with varying bag constant. Our calculation also shows 
that the mixed hybrid star has radius (for the maximum mass) quite different from
the neutron or strange star, their radius lying between the neutron and strange star.

Observationally, the surface magnetic field of most of the pulsars are in the ranges of $10^8$ to 
$10^{12}$G. Such fields have almost no effect in the EOS of matter in those stars. However, for 
magnetars the magnetic fields are very high ($\sim\ 10^{17}-10^{18}$G). Flux conservation from
progenitor stars can give rise to magnetic field of field strength $2-3$ orders higher. The 
mass-radius relationship for a mixed hybrid star is quite different from the pure neutron or strange star,
and so it is likely to have different observational characteristics. It is also clear that
magnetars are different from normal pulsars, as they have lesser mass due to flatter 
EOS. It is to be mentioned here that we have only considered effect from Landau quantization and found 
that they have significant effect on the mixed phase region once it is greater than $10^{14}$G.
Here we have not considered the effect from anomalous magnetic moment.
Anomalous magnetic moment stiffens the EOS, but their
effect is significant if the magnetic field strength is of the order of  $10^{19}$G. For such high
magnetic fields the NS becomes unstable. Therefore our consideration of only the effect from
Landau quantization seems alright. 
As the interiors of the compact stars are hidden from direct observation, we have to rely only on the
observations coming from their surface. Recent developments has been made on measuring accurately 
the mass of compact stars but a exact measurement of their radius is still not possible \cite{Demorest10}. 
The knowledge of the radius of a compact stars can really give us the hint of the matter 
components at the star interiors, as
we have seen here that different EOS provide different mass-radius relationship. 

{}
\end{document}